\documentclass[aps,twocolumn,prb,superscriptaddress,floats,nobibnotes,notitlepage,nofootinbib]{revtex4-2}

\usepackage{graphicx}
\usepackage[utf8]{inputenc}
\usepackage{amsmath,amssymb}
\usepackage{newtxtext}
\usepackage[varvw]{newtxmath}
\usepackage{multirow}
\usepackage{mathtools}
\usepackage{enumerate}
\usepackage[pdftex]{hyperref}
\usepackage[normalem]{ulem}

\hypersetup{
    colorlinks,
    linkcolor={blue},
    citecolor={blue},
    urlcolor={blue}
}

\def\<{\langle}
\def\>{\rangle}

\DeclareMathOperator{\sgn}{sgn}

\renewcommand{\Im}[0]{\text{Im}\,}

\newcommand{\ve}[1]{\boldsymbol{#1}}
\newcommand{\mf}{\text{MF}}
\newcommand{\Hlin}{H_{1}}
\newcommand{\Hquad}{H_{2}}
\newcommand{\Jk}{J_{\mathrm{K}}}
\newcommand{\Jkc}{J_{\mathrm{K}}^{\mathrm{c}}}
\newcommand{\Jh}{J_{\mathrm{H}}}

\makeatother

\makeindex

\date{\today}

\begin{document}

\title{Magnetic  quantum phase  transition in a  metallic Kondo  heterostructure}

\author{Zi Hong Liu}
\affiliation{Institut f\"ur Theoretische Physik und Astrophysik and W\"urzburg-Dresden Cluster of Excellence ct.qmat,\\
Universit\"at W\"urzburg, 97074 W\"urzburg, Germany}
\author{Bernhard Frank}
\affiliation{Institut f\"ur Theoretische Physik and W\"urzburg-Dresden Cluster of Excellence ct.qmat,
Technische Universit\"at Dresden, 01062 Dresden, Germany}
\author{Lukas Janssen}
\affiliation{Institut f\"ur Theoretische Physik and W\"urzburg-Dresden Cluster of Excellence ct.qmat,
Technische Universit\"at Dresden, 01062 Dresden, Germany}
\author{Matthias Vojta}
\affiliation{Institut f\"ur Theoretische Physik and W\"urzburg-Dresden Cluster of Excellence ct.qmat,
Technische Universit\"at Dresden, 01062 Dresden, Germany}
\author{Fakher F. Assaad}
\affiliation{Institut f\"ur Theoretische Physik und Astrophysik and W\"urzburg-Dresden Cluster of Excellence ct.qmat,\\
Universit\"at W\"urzburg, 97074 W\"urzburg, Germany}

\begin{abstract}
We   consider a  two-dimensional    quantum  spin  system  described  by  a  Heisenberg  model that is  embedded  in a  three-dimensional   metal.  
The  two  systems  couple    via  an   antiferromagnetic  Kondo  interaction.  In such  a  setup,  the  ground  state  generically remains  metallic   down  to the lowest  temperatures  and  allows us to  study  magnetic  quantum phase  transitions  in metallic  environments.    From the  symmetry point of  view,  translation  symmetry is present in  two out of  three lattice  directions  such  that
 crystal momentum is only partially  conserved.  Importantly,  the  construction   provides  a  route  to   study,  with  negative-sign-free auxiliary-field  quantum Monte   Carlo  methods,   the physics of  local moments in metallic  environments. Our  large-scale numerical  simulations  show  that  as  a  function of  the Kondo coupling,  the system  has  two  metallic  phases.  In the limit of strong Kondo coupling,    a paramagnetic  heavy-fermion  phase   emerges.   Here, the spin  degree  of  freedom is  screened  by  means of  the   formation of  a   composite quasiparticle  that  participates in the Luttinger  count.    At  weak  Kondo coupling,   magnetic order  
 is  present.   This phase  is  characterized by  Landau-damped  Goldstone modes.  Furthermore,  the  aforementioned  composite  quasiparticle  remains  
 \textit{intact}   across   the quantum phase  transition.   
\end{abstract}

\maketitle

%%%%%%%%%%%%%%%%%%%%%%%%%%%%%%%%%%%%%%%%%%%%%%%%%%%%%%%%%%%%%%%%%%%%%%%

\section{Introduction}

The  interplay  of  quantum spins  with itinerant   electrons  is  pivotal  for  the 
understanding  of   heavy-fermion systems  \cite{Lohneysen_rev,Coleman07_rev,Tsunetsugu97_rev} as  well as  for  high-temperature  superconductivity \cite{Lee06_rev,Emery87,Zhang88}. 
Magnetic  heterostructures, in  which  the  magnetic  system  has a lower  dimension as  the metallic host,    provide  a  particularly   rich  realization of the above.     For  example,   Heisenberg  spin chains  on metallic   surfaces  
 with  a  Kondo  coupling  between  the  spins  and  conduction electrons   provide  experimental  realizations of such dimensional  mismatch \cite{Toskovic16}.     Interestingly, many phases  and  phase  
transitions  can  be realized in this  setup.  At  large  Kondo  coupling,   the spins  are  Kondo screened, leading to the  formation of a  composite   quasiparticle that  participates  in the Luttinger  volume \cite{Danu21,Raczkowski22}.     At  weak  couplings,   the  fate of the spin chain   depends  on    the   nature of the two-dimensional Fermi  surface.  For  Dirac  systems  with  point-like  Fermi surfaces, 
 the  Kondo  exchange  is  irrelevant   at  weak    coupling,     thus leading to an  FL*  phase  and  the   absence  of the  aforementioned  emergent  composite    fermion  \cite{Danu20}.   In contrast,  for  a  generic  Fermi  surface,   one  observes   dissipation-induced order  and  
 no  destruction of the composite fermion  
 across the phase  transition \cite{Danu22}.   Hence,  depending on the  nature of the metallic  state,  one observes  Kondo breakdown or 
  Hertz-Millis-type  transitions.

 	The  aim  of  this  article  is  to  generalize  the  above  to  two-dimensional  spin systems in a  three-dimensional  metallic  environment.    This  choice of dimensions   relates, for instance,   to  heterostructures  of   CeIn$_3$    monolayers  embedded  in      LaIn$_3$  \cite{shishido10} in the  limit   where the  interlayer    distance  is  large.     We  will show  that  as  a function of  the Kondo interaction,   our  model  system   undergoes a  magnetic  quantum phase  transition.  Both  phases  are  metallic.  In the  strong-coupling limit,  we observe  a  heavy-fermion  metallic  state,  and  the emergence  of a composite  fermion   that  participates in the Luttinger  count.    To  make  this statement  precise,  we  note  that our model   is  invariant  under  translations along  the magnetic  plane,  but  not perpendicular to it.  Hence,   we  can    understand it as  a multi-band model,   where  the  direction perpendicular   to the magnetic plane  reflects  the band index.   Within  this setup,  the  paramagnetic heavy-fermion phase   has $L+1$ electrons  per  unit  cell corresponding to $L$ conduction electrons and   the local-moment electron.  
	The  magnetic phase  is  characterized by   Landau-damped  magnons.  In this phase,  we  show  that the  composite fermion  remains intact in the sense  that  it participates in the Luttinger count.   The numerical and  analytical  data  we will  present in this  article aims  at documenting the  interpretation that the  model  provides  a  Hertz-Millis-type  transition  for  quantum spins  embedded in a metallic  environment.

From  the  technical point  of  view    magnetic  heterostructures  are  particularly  appealing  since  they  provide a route  to  
\textit{beat} the   infamous  sign problem  in  quantum Monte Carlo (QMC)   simulations.    In particular,  for  systems   with an  extensive 
 number of  quantum  spins  and in  dimensions  greater than unity, metallic  states invariably  suffer  from  the  negative-sign problem. 
 This  stems  from  the  fact  that  for   repulsive  interactions,  necessary  for the  very 
 formation of  local moments,  the condition for   the  absence of the  negative-sign problem  requires   
 particle-hole   symmetry \cite{Hirsch85,Wu04,Li16}.   This results in a nested  Fermi  surface,  such that at  low  enough  temperatures  
 a magnetic insulating state  will  occur.  In contrast  a sub-extensive   amount of local moments  will  not be  able  to gap  out  the  Fermi  surface.

	The  rest of this work  is organized as follows. In Sec.~\ref{sec:Model}, we introduce the Hamiltonian and the lattice construction of the microscopic model. Furthermore, we discuss the nature of the model in the weak-interaction limit and in the large-$N$ limit. In Sec.~\ref{sec:QMC}, we describe our implementation of the QMC approach for the present context. In Sec.~\ref{sec:QMC_result}, we present our unbiased numerical results. Based on these results, we discuss the dynamical behavior of local spin and fermion excitations. Our conclusions and an outlook are given in Sec.~\ref{sec:Conclusion}.
	
%%%%%%%%%%%%%%%%%%%%%%%%%%%%%%%%%%%%%%%%%%%%%%%%%%%%%%%%%%%%%%%%%%%%%%%

\section{Model}\label{sec:Model}

\begin{figure}[tb]
\begin{centering}
\includegraphics[width=\linewidth]{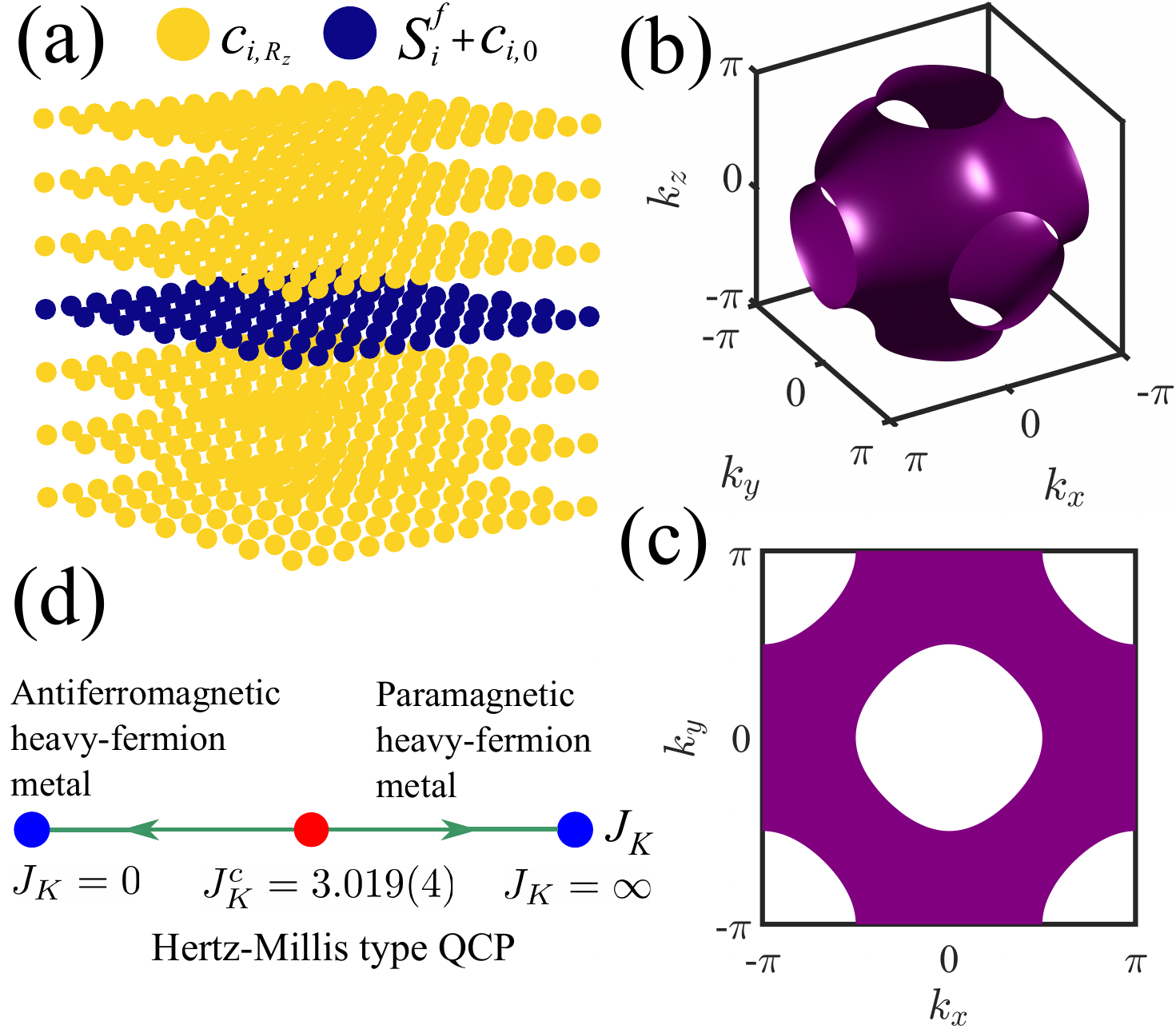}
\par\end{centering}
\caption{(a) Sketch of Kondo heterostructure, consisting of a two-dimensional  array  of  magnetic impurities (blue dots) and three-dimensional  itinerant conduction electrons,  modeled by  a tight-binding Hamiltonian on a  cubic lattice (yellow dots). (b)  Three-dimensional Fermi surface of    conduction electrons.  (c) Projected Fermi surface. (d) Ground-state phase diagram of model in Eq.~\eqref{eq:model}, as extracted from QMC results.}
\label{fig:plot_of_model}
\end{figure}

We propose a model of a Kondo heterostructure in  which  a layer of magnetic impurities  is  embedded  in a   three-dimensional  metal  as  depicted 
in   Fig.~\ref{fig:plot_of_model}(a).     The  metallic  environment is modeled by  a  tight-binding Hamiltonian  on a   cubic lattice  of linear length $L$ and with translation  invariance  in the  $x$, $y$, and $z$ directions. For the magnetic  layer, we employ a  Heisenberg  model  with  exchange  $J_\mathrm{H}$  on a  square  lattice  with the same lattice constant  as  that of
the three-dimensional cubic lattice.      The  two  subsystems  are coupled via  a  Kondo interaction  $J_\mathrm{K}$.
Specifically,  the Hamiltonian for this Kondo-lattice-model heterostructure (KLM-hetero) is defined as
\begin{equation}
\hat{H}_{\text{KLM-hetero}} =\hat{H}_{\text{Fermi}}+\hat{H}_{\text{Heisenberg}}+\hat{H}_{\text{Kondo}}.
\label{eq:model}
\end{equation}
Here, 
\begin{equation}
\hat{H}_{\text{Heisenberg}}=\Jh\sum_{\left\langle \ve{i},\ve{j} \right\rangle }\hat{\boldsymbol{S}}_{\ve{i}}^{f}\cdot \hat{\boldsymbol{S}}_{\ve{j}}^{f}
\end{equation}
describes antiferromagnetic  spin-1/2   Heisenberg  interactions on  nearest-neighbor bonds $\left\langle \ve{i},\ve{j} \right\rangle$   of the  square  lattice.  
The Hamiltonian of the three-dimensional metal reads
\begin{align}\label{eq:HFermi}
\hat{H}_\mathrm{Fermi} & =-t\sum_{\left\langle (\ve{i},R_{z}),(\ve{j},R'_{z})\right\rangle ,\sigma}\left(\hat{c}_{\ve{i},R_{z},\sigma}^{\dagger}\hat{c}_{\ve{j},R'_{z},\sigma}^{\phantom{\dagger}}+\text{h.c.}\right)\nonumber \\
 & =\sum_{\boldsymbol{k}_{2},k_{z},\sigma}\epsilon_{\boldsymbol{k}_{2},k_{z}}\hat{c}_{\boldsymbol{k}_{2},k_{z},\sigma}^{\dagger}\hat{c}_{\boldsymbol{k}_{2},k_{z},\sigma}^{\phantom{\dagger}}.
\end{align}
Here, $\hat{c}_{\ve{i},R_z,\sigma}^{\dagger}$   creates  an  electron  with  $z$-component of spin  $\sigma$   in a Wannier  state  centered  around the lattice  site $(\ve i,R_z)$  of the  cubic  lattice, and hopping on nearest-neighbor bonds  $\left\langle (\ve i,R_z),(\ve j,R_z') \right\rangle$ in all three directions. 
 
We  use   periodic   boundary  conditions,    and  define  Bloch  states, 
\begin{equation}
   \hat{c}^{\dagger}_{\ve{k}_2,k_z}    =  \frac{1}{\sqrt{L^3}} \sum_{\ve{i},R_z}  e^{i  (\ve{k}_2\cdot \ve{i}  +  k_z R_z) }  \hat{c}_{i,R_z,\sigma}^{\dagger} 
\end{equation}
with three-dimensional  crystal momentum $\boldsymbol{k}=(\boldsymbol{k}_2, k_z)   \equiv (k_x,k_y,k_z)$. The dispersion  relation    reads  $\epsilon_{\boldsymbol{k}_{2},k_{z}}=-2t(\cos k_{x} + \cos k_{y} + \cos k_{z} )$,  and in the absence  of  coupling to the magnetic plane,    the   three-dimensional crystal momentum  is conserved  up to a  reciprocal lattice  vector.   The  Fermi   surface  of   the  metal is  shown in Fig.~\ref{fig:plot_of_model}(b).

$\hat{H}_{\text{Kondo}}$ describes the Kondo coupling between the $c$ conduction electrons and the magnetic impurities, 
\begin{equation}\label{eq:HKondo}
\hat{H}_{\text{Kondo}}=\Jk\sum_{i}\hat{\boldsymbol{S}}_{\ve{i},R_{z}=0}^{c}\cdot \hat{\boldsymbol{S}}_{\ve{i}}^{f},
\end{equation}
with coupling strength $\Jk$   and  $\hat{\boldsymbol{S}}_{\ve{i},R_{z}}^{c} = $ 
$ \frac{1}{2} \sum_{\sigma,\sigma'}\hat{c}_{\ve{i},R_{z},\sigma}^{\dagger}   \ve{\sigma}_{\sigma,\sigma'} \hat{c}_{\ve{i},R_{z},\sigma'}^{\phantom{\dagger}}$.  Importantly,   the  two-dimensional  array of  magnetic impurities couples  to  the  layer of  conduction electrons  at  $R_{z} = 0$, such  that   $k_z$ is  no longer a  good  quantum number.      
Low-energy scattering processes then involve states on the projected Fermi surface, obtained from the summation over all $k_z$.
Technically, the  projected  Fermi surface  can be defined as  the support of
\begin{equation}
A_{c,0}^{R_{z}}\left(\boldsymbol{k}_{2},\omega=0\right)=-\frac{1}{\pi}\mathrm{Im}\left\{ G_{c,0}^{R_{z}R_{z}}\left(\boldsymbol{k}_{2},\omega=0\right)\right\},
\end{equation} 
where 
\begin{multline}
G_{c,0}^{R_{z}R_{z}'}(\ve{k}_2,\omega) = -i \sum_{i,\sigma}  \int_0^\infty dt\, e^{i \ve{k}_2 \cdot \ve{r}_i + i \omega t}
\\
\langle \{{\hat c}^{\phantom{\dagger}}_{i,R_z,\sigma}(t), {\hat c}^\dagger_{0,R_z',\sigma}(0) \}\rangle^{\phantom{\dagger}}_0
\end{multline}
denotes the noninteracting electronic Green's function in the two-dimensional reciprocal space.
The projected Fermi surface is depicted in  Fig.~\ref{fig:plot_of_model}(c).

\subsection{Weak-coupling limit}
\label{Weak_coupling:sec}
At  $\Jk =0$, spins and conduction electrons  decouple.    To set the stage, we  will first  discuss  these degrees of  freedom  separately,  and   then  
investigate  how  they  couple perturbatively  in  $\Jk$.   

 The  spin and  charge  excitations  of  the  conduction  electrons  are  characterized 
by  the    noninteracting susceptibility  
\begin{equation}
	    \chi^{0}(\ve{r}  -   \ve{r}',  \tau -  \tau')     \equiv   \frac{1}{4} \langle \hat{\ve{c}}^{\dagger}_{\ve{r}}(\tau) \ve{\sigma}  \hat{\ve{c}}^{\phantom\dagger}_{\ve{r}}(\tau)   
	    \cdot  \hat{ \ve{c}} ^{\dagger}_{\ve{r}'}(\tau') \ve{\sigma}   \hat{\ve{c}} ^{\phantom\dagger}_{\ve{r}'}(\tau')  \rangle_0
\end{equation}
with $\hat{\ve{c}}_{\ve{r}} = (c_{\ve{r},\uparrow}, c_{\ve{r},\downarrow})$, and
where the  expectation value is taken  with respect  to  $  \hat{H}_\text{Fermi}$.   To  simplify  the notation, we set 
$\ve{r} = (\ve{i}, R_z) $.
The   particle-hole   symmetric   conduction  band   remains  invariant  under the  transformation  $   \hat{c}^{\dagger}_{\ve{r}}   \rightarrow 
  e^{i \ve{Q} \cdot  \ve{r} }\hat{c}^{\phantom\dagger}_{\ve{r}}$ with $\ve{Q}=\left(\pi,\pi,\pi \right)$, 
such  that
  \begin{equation}
  \chi^{0}( \ve{r}, \tau )     = \frac{3}{2} e^{i \ve{Q} \cdot  \ve{r} } 
  \left[  \frac{1}{\left( 2 \pi\right)^3} \int_{\text{BZ}} d^{3}\ve{k}  \, e^{i \ve{k}\cdot \ve{r}}   e^{\tau \epsilon(\ve{k})}  f(\epsilon(\ve{k})) 
  \right]^2,
  \end{equation}  
where  $\epsilon(\ve{k})  =  -2t  (   \cos k_x  +    \cos k_y +   \cos k_z )  $,  $f(\epsilon) $  is  the  Fermi function,    and  we  have  set the lattice  constant to unity.      From the above,   one  will see  that  at   zero  temperature,  
  \begin{equation}
  	   \chi^{0}( \ve{0}, \tau )     =   \frac{3}{2}
  \left(  \int_{-\infty}^{0} d \epsilon  \, e^{\tau \epsilon}    N(\epsilon)     \right)^2 
  \label{eq:chi0_c}
  \end{equation}
 where  $ N(\epsilon)  =  \frac{1}{\left( 2 \pi\right)^3} \int_{\text{BZ}} d^{3}\ve{k}  \, \delta\left(\epsilon -  \epsilon(\ve{k}) \right)  $  is  the density of  states. Since  in  three dimensions $N(\epsilon)$  has  no  singularity  at   the   Fermi  energy,  $\epsilon_\mathrm{F}=0$, and  since  the long-imaginary-time  behavior   of  the integral    
 stems    from energies  close  to the Fermi  surface,  we  obtain the  asymptotic form  $  \chi^{0}( \ve{0}, \tau )  \sim   1/\tau^2$ for large $\tau$.   We  now   consider  the spatial  decay  at  equal time.  For  a  spherical  Fermi  surface,    the   $\ve{k}$ integration  can be computed  exactly  to  obtain the   large-distance  behavior $  \chi^{0}( \ve{r}, 0)  \sim   1/|\ve{r}|^4 $.

For  the  Heisenberg  model,   we  follow   Haldane's derivation of  the  O(3)  non-linear  sigma  model \cite{Haldane88}.  
The  starting  point  is  a  spin-coherent-state   formulation    of  the path  integral.   In  the   large-$S$  limit and  assuming   dominant  antiferromagnetic    spin-spin  fluctuations,    the  action for the local moments reads  
 \begin{equation}
    S_\text{AFM}   =      \rho_s \int d^2 \ve{x}    d \tau   \,      \sum_{\mu} \left( \partial_{\mu} \ve{\Omega}(\ve{x},\tau) \right)^2.	   
 \end{equation}
In the above,  we  have  neglected  the  Berry phase   since it  plays  no  dominant  role  in the  ordered  state,  $\ve{\Omega}(\ve{x},\tau)$   is  a    space-time-dependent   unit  vector   accounting  for    the  dynamics  of the  antiferromagnetic   O(3)   order  parameter,   and  $\mu$   runs  over  the  temporal  and    spatial  directions.   Finally,   we   have  set  the  spin wave  velocity  to   unity.    In the   ordered  state,   the O(3)  symmetry is  reduced to O(2),   and   for  spontaneous   symmetry  breaking  along  the $z$ direction   we   consider  the  ansatz
\begin{equation}
\label{sdw_ansatz.eq}
	 \ve{\Omega}(\ve{x},\tau)   =  \left(  \ve{n}(\ve{x},\tau),   \sqrt{1  -  \ve{n}^2(\ve{x},\tau) } \right)    
\end{equation}
where $\ve{n}=(n_x,n_y)$ denotes the transverse components of the order parameter,
with  $ |\ve{n}|  \ll 1 $.   Expanding  in    $ |\ve{n}|  $  gives,   $S_\text{AFM} =  S^{(0)}_\text{AFM} +   S^{(1)}_\text{AFM} +  \cdots$     with  
$   S^{(0)}_\text{AFM} =   \rho_s \int d^2 \ve{x}    d \tau   \,      \sum_{\mu} \left( \partial_{\mu} \ve{n}(\ve{x},\tau) \right)^2   $  and 
$ S^{(1)}_\text{AFM}    = \frac{\rho_s}{4} \int d^2 \ve{x}    d \tau   \,      \sum_{\mu} \left( \partial_{\mu} \ve{n}^2(\ve{x},\tau) \right)^2    $.     Under  the  scale  transformation $\ve{x}  \rightarrow \lambda \ve{x}  $,  $\tau  \rightarrow \lambda \tau  $   and  $\ve{n} \rightarrow   \frac{\ve{n}}{\sqrt{\lambda}} $,  $S^{(0)}_\text{AFM}$   remains  invariant   and  accounts for   the   Lorentz-symmetric gapless  transverse   spin-spin   fluctuations.    Under this  transformation,  
the magnon-magnon     interactions   described  by    $S^{(1)}_\text{AFM}$   scale  as   $1/\lambda$  and  are  hence  irrelevant  at     the  \textit{spin-wave}  fixed point.  Higher  orders in the  expansion are  even more irrelevant.

With this background,  we   can now  couple the   two  systems perturbatively.  Using  fermion-coherent  states  for the conduction electrons  and   the  spin-coherent  states  for   the  local  moments,   the partition function maps  onto  a   bilinear  fermionic  problem   interacting   with  the space-and-time-dependent   spin-coherent state.   At  this point,  one  can integrate out the fermions and  expand  the  resulting  action up to second order in    the Kondo coupling $\Jk$.    Omitting  the  Berry  phase,    the  resulting  action reads
\begin{multline}
	   S    =   S_\text{AFM} +  \Gamma  \int d^2 \ve{x} d^2 \ve{x}'   d \tau  d  \tau'   \\
	      \ve{\Omega} (\ve{x},\tau)  \chi^{0}_\text{AFM} (\ve{x}-\ve{x}', \tau - \tau'  ) \ve{\Omega} (\ve{x}',\tau'). 
\end{multline}
Here,   $  \chi^{0}_\text{AFM} (\ve{x}, \tau ) $  $=    \frac{3}{2} 
  \left[  \frac{1}{\left( 2 \pi\right)^3} \int_{\text{BZ}} d^{3}\ve{k}  \, e^{i \ve{k}_2 \cdot \ve{x}}   e^{\tau \epsilon(\ve{k})}  f(\epsilon(\ve{k}))  \right]^2  $
 and   $\Gamma  \propto  \Jk^2$.     Note  that   the  magnetic layer  lies at  $R_z=0$ and  $\ve{k}  =  (\ve{k}_2,k_z) $.
 
 Let us     concentrate on the  equal  time  spacial  and   local  temporal correlations.   In  this case,   
 \begin{align}
	   S    & \simeq S_\text{AFM}   +   \Gamma  \int d^2 \ve{x}  d \tau  d  \tau'   
	         \frac{  \ve{\Omega} (\ve{x},\tau) \cdot \ve{\Omega} (\ve{x},\tau')}{(\tau - \tau')^2}  
	         \nonumber   \\ & \quad
	          + \Gamma   \int  d^{2} \ve{x} d^{2}\ve{x}'  d\tau   \frac{  \ve{\Omega} (\ve{x},\tau) \cdot \ve{\Omega} (\ve{x}',\tau)}{| \ve{x} -  \ve{x}' |^4}.  
\end{align}
The  ansatz of   Eq.~\eqref{sdw_ansatz.eq}      then gives     $S    =   S^{(0)}   +  S^{(1)}   +    \cdots  $  with 
\begin{align}
\label{S_landau_damped.eq}
    S^{(0)}   &   =  \rho_s \int d^2 \ve{x}    d \tau  \left[ \left( \partial_{x} \ve{n}(\ve{x},\tau) \right)^2    +  \left( \partial_{y} \ve{n}(\ve{x},\tau) \right)^2 \right]     
    \nonumber \\
        &\quad + \Gamma  \int d^2 \ve{x}  d \tau  d  \tau'    \frac{  \ve{n} (\ve{x},\tau) \cdot \ve{n} (\ve{x},\tau')}{(\tau - \tau')^2} 
     \nonumber  \\ 
       &\quad +    \Gamma  \int d^2 \ve{x}  d^2 \ve{x}'  d  \tau   \frac{  \ve{n} (\ve{x},\tau) \cdot \ve{n} (\ve{x}',\tau)}{| \ve{x} -  \ve{x}'|^4} 
 \end{align}
 and 
 \begin{align}
    S^{(1)}   & = \rho_s \int d^2 \ve{x}    d \tau \left[ \left( \partial_{\tau} \ve{n}(\ve{x},\tau) \right)^2     +    
     \frac{1}{4}\sum_{\mu} \left( \partial_{\mu} \ve{n}^2(\ve{x},\tau) \right)^2   \right]
     \nonumber    \\  
     &\quad + \frac{\Gamma}{4}  \int d^2 \ve{x}  d \tau  d  \tau'  \,  \frac{  \ve{n}^2 (\ve{x},\tau) \ve{n}^2 (\ve{x},\tau')}{(\tau - \tau')^2}  
     \nonumber    \\ 
     &\quad  + \frac{\Gamma}{4}  \int d^2 \ve{x}  d^2 \ve{x}'  d  \tau  \,  \frac{  \ve{n}^2 (\ve{x},\tau) \ve{n}^2 (\ve{x}',\tau)}{|\ve{x} - \ve{x'} |^4}.  
 \end{align}
Under  the    scale  transformation    $  \ve{x}   \rightarrow  \lambda \ve{x} $,   $  \tau   \rightarrow  \lambda^2 \tau $    and   
$\ve{n}  \rightarrow \frac{\ve{n} } {\lambda}  $,  $ S^{(0)}$   remains  scale  invariant  and  describes a Landau-damped  Goldstone-mode
fixed  point  with   dynamical  exponent $z=2$.   At  this  fixed  point,   $ S^{(1)} $  is  irrelevant.  

To conclude,   the   dynamical  spin-structure  factor in the weak-coupling limit  is  expected  to show  long-range magnetic order   
and to be  described  by  Landau-damped   Goldstone   modes  governed  by  the  fixed-point action of  
Eq.~\eqref{S_landau_damped.eq}.       A corresponding spin-wave  analysis,  presented in  Appendix~\ref{spin_wave.app},  confirms this point  of  view.

\subsection{Mean-field approximation}\label{sec:MF}
In this section, we   consider   a  mean-field  approximation  that   accounts  for   Kondo screening as  well as  for  magnetic  ordering \cite{Zhang00b}. 
 We  use  the pseudo-fermion representation  $\ve{\hat{f}}_{\ve{i},\sigma}^{\dagger} =  \left( \hat{f}^{\dagger}_{\ve{i},\uparrow},\hat{f}^{\dagger}_{\ve{i},\downarrow}\right)  $   of the spin-1/2, $\hat{\boldsymbol{S}}_{\ve{i}}^{f}=\frac{1}{2}\ve{\hat{f}}^{\dagger}_{\ve{i}}\boldsymbol{\sigma}\ve{\hat{f}}^{\dagger}_{\ve{i}} $,   that  holds  provided   that    we  
 impose  the  constraint $\hat{Q}_{\ve{i}}= \ve{\hat{f}}^{\dagger}_{\ve{i}} \ve{\hat{f}}^{\phantom\dagger}_{\ve{i}} =1 $.
With  this  choice,  the  Kondo  coupling can be  written as  
\begin{align}
\boldsymbol{\hat{S}}_{\ve{i}, R_{z}=0}^{c}\cdot\boldsymbol{\hat{S}}_{\ve{i}}^{f} & = \hat{S}_{\ve{i}, R_{z}=0}^{c,z}\hat{S}_{\ve{i}}^{f,z}
 \nonumber \\ & \quad
 - \frac{1}{4}\sum_{\sigma}\left(\hat{c}_{\ve{i},R_{z}=0,\sigma}^{\dagger}\hat{f}^{\phantom\dagger}_{\ve{i},\sigma}+\hat{f}_{\ve{i},-\sigma}^{\dagger}\hat{c}_{\ve{i},R_{z}=0,-\sigma}^{\phantom\dagger}\right)^{2},
\label{eq:kondo_term}
\end{align}
where $\hat{S}_{\ve{i}, R_z=0}^{c,z}=\frac{1}{2}\hat{\ve{c}}_{\ve{i},R_z=0}^{\dagger}\sigma_z\hat{\ve{c}}_{\ve{i},R_z=0}^{\phantom\dagger}$.  

 The  above  reformulation  allows  us  to  carry  out  mean-field  approximations  that  account  for the  Kondo effect, as  described in the large-$N$ limit,  and  magnetism.     We  note   that this mean-field   decomposition  can  be  formulated  for   an SU($N$)-symmetric  Kondo lattice  model \cite{Raczkowski20},   in which    magnetism   driven  by the RKKY  interaction  becomes  a  $1/N$  effect. 
   In the  mean-field  approximation,   squared order-parameter fluctuations  are  neglected,   i.e.,  
 \begin{align}
\hat{O}^{2} & =\left(\left\langle \hat{O}\right\rangle +\Delta\hat{O}\right)^{2}=\left\langle \hat{O}\right\rangle ^{2}+2\left\langle \hat{O}\right\rangle \Delta\hat{O}+\left(\Delta\hat{O}\right)^{2}\nonumber \\
 & \approx2\left\langle \hat{O}\right\rangle \hat{O}-\left\langle \hat{O}\right\rangle ^{2},
\label{eq:hf_approx}
\end{align}
where $\Delta\hat{O}=\hat{O}-\left\langle \hat{O}\right\rangle$  are  the fluctuations.   Here,  we  consider the following order parameters 
\begin{equation}
\begin{cases}
\left\langle \hat{S}_{\ve{i}, R_{z}=0}^{c,z}\right\rangle =-m_{c}e^{i\boldsymbol{Q} \cdot\boldsymbol{i} }, \\
\left\langle \hat{S}_{\ve{i}}^{f,z}\right\rangle =m_{f}e^{i\boldsymbol{Q} \cdot\boldsymbol{i}}, \\
\left\langle \hat{c}_{\ve{i},R_z=0,\sigma}^{\dagger}\hat{f}_{\ve{i}, \sigma}^{\phantom\dagger}+\hat{f}_{\ve{i}, -\sigma}^{\dagger}\hat{c}_{\ve{i}, R_z=0,-\sigma}^{\phantom\dagger}\right\rangle=V, 
\end{cases}
\label{eq:mft_order}
\end{equation}
where $V$ accounts for the hybridization between  the  $c$ and $f$ fermions,   $m_c$ and $m_f$  denotes  the magnetizations arising from conduction electrons and  impurity  spins,  respectively and  $\ve{Q} = (\pi,\pi) $.   By combining Eqs.~\eqref{eq:kondo_term}-\eqref{eq:mft_order}, we obtain the effective mean-field Hamiltonian
\begin{align}
\hat{H}_{\text{MF}} & = \hat{H}_\mathrm{Fermi}+\sum_{\ve{i}}\lambda_{\ve{i}}\left(\hat{f}_{\ve{i},\sigma}^{\dagger}\hat{f}_{\ve{i},\sigma}^{\phantom\dagger}-1\right)
\nonumber \\ & \quad 
-\frac{\Jk V}{2}\sum_{\ve{i},\sigma}\left(\hat{c}_{\ve{i},R_z=0,\sigma}^{\dagger}\hat{f}_{\ve{i},\sigma}^{\phantom\dagger}+\text{h.c.}\right)
\nonumber \\ & \quad 
+ \Jk \sum_{\ve{i} }e^{i\boldsymbol{Q} \cdot \ve{i}}\left(m_{f}\hat{S}_{\ve{i}, R_{z}=0}^{c,z}-m_{c}\hat{S}_{\ve{i}}^{f,z}\right)
\nonumber \\ & \quad 
- \Jh m_{f}  z_c \sum_{\ve{i}}e^{i\boldsymbol{Q}\cdot\boldsymbol{i}}\hat{S}_{\ve{i}}^{f,z}
+\epsilon_{0}
 \label{eq:mft_ham}
\end{align}
where $\epsilon_{0}=L^2\left[2\Jh\left(m_{f}\right)^{2}+\Jk m_{c}m_{f}+ \Jk V^{2}/2\right]$   and  $z_\mathrm{c}=4$ corresponds  to the coordination number
of  the  square lattice. To suppress the charge fluctuations in the pseudo fermion sector, we introduce the Lagrange parameter ${\lambda_{\ve{i}}}$ in the first line of Eq.~\eqref{eq:mft_ham}, which imposes the constraint $\hat{Q}_{\ve{i}}=1$ at each site. 
 
In the  above,  we  have  not accounted  for  a  \textit{spinon}  description of  the  quantum antiferromagnet,   in which  the   pseudo fermions  delocalize 
in the  magnetic  impurity  plane.      In the magnetic  phase,  where   the hybridization matrix element  vanishes,  we  can justify this choice from 
our  knowledge  that  the  two-dimensional   Heisenberg model on the  square lattice  does  not  have a  fractionalized ground  state.   In the heavy-fermion  state,  $V \neq  0$,    the  $f$ pseudo fermions  acquire   electric  charge and lose   their  gauge  charge  via   the  Higgs  mechanism, 
 such  that  they  can  acquire  a   dispersion relation \cite{Danu21,Raczkowski22}.
The mean-field Hamiltonian $\hat{H}_{\text{MF}}$ is bilinear  in the fermions  and  can hence be solved numerically exactly in polynomial time  for a given  set  of  order parameters.   For an  analytical   calculation  in the  heavy-fermion state,  we  refer to Appendix~\ref{mean_field.app}.  
The  order parameters  are obtained  by the minimizing   the free energy $\partial F_{\text{MF}}/\partial m_{c} = \partial F_{\text{MF}}/\partial m_{f}=\partial F_{\text{MF}}/\partial V=0$,   leading to a set of self-consistent equations
\begin{equation}
\begin{cases}
m_{c}=-\frac{1}{2L^2}\sum_{\ve{k}_{2},\sigma}\left(-1\right)^{\sigma}\left\langle \hat{c}_{\ve{k}_{2}+\ve{Q},R_{z}=0,\sigma}^{\dagger}\hat{c}_{\ve{k}_{2},R_{z}=0,\sigma}^{\phantom\dagger}\right\rangle, \\
m_{f}=\frac{1}{2L^2}\sum_{\ve{k}_{2},\sigma}\left(\left(-1\right)^{\sigma}\left\langle \hat{f}_{\ve{k}_{2}+\ve{Q},\sigma}^{\dagger}
\hat{f}_{\ve{k}_{2},\sigma}^{\phantom\dagger} \right\rangle \right), \\
V=\frac{1}{2L^2}\sum_{\ve{k}_{2},\sigma}\left(\left\langle \hat{c}_{\ve{k}_{2},R_{z}=0,\sigma}^{\dagger}\hat{f}_{\ve{k}_{2},\sigma}^{\phantom\dagger}
+\hat{f}_{\ve{k}_{2},\sigma}^{\dagger}\hat{c}_{\ve{k}_2,R_{z}=0,\sigma}^{\phantom\dagger}\right\rangle \right), \\
\frac{1}{L^2}\sum_{\ve{i},\sigma}\left\langle \hat{f}_{\ve{i},\sigma}^{\dagger}\hat{f}_{\ve{i},\sigma}^{\phantom\dagger}\right\rangle =1.
\end{cases}
\label{eq:mft_selfconsist}
\end{equation}
The last equation in Eq.~\eqref{eq:mft_selfconsist}  corresponds  to  the half-filling constraint for the  pseudo fermions.  At $\lambda =0$,  the  mean-field Hamiltonian  of Eq.~\eqref{eq:mft_ham}  is particle-hole symmetric  such  that this choice  of  the  Lagrange parameter  satisfies  the  constraint on 
average. 
Technical details concerning the numerical solution of the self-consistency equations are provided in Appendix~\ref{self_consistency.app}.

\begin{figure}[tb]
\begin{centering}
\includegraphics[width=0.9\linewidth]{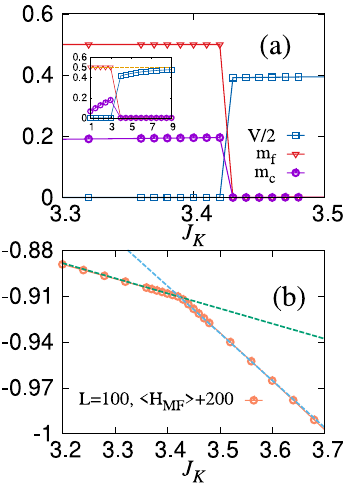}
\par\end{centering}
\caption{
%Ground-state phase diagram of mean-field Hamiltonian $\hat{H}_{\text{MF}}$. We have set $t=1$, $\Jh=0.5$, and used a linear system $L=150$. Dashed lines separate antiferromagnetic heavy-fermion (AFM), coexistence (Coexist), and paramagnetic heavy-fermion (PM) phases.
(a) Mean-field order parameters  at zero temperature. Here,  $t=1$, $\Jh=0.5$, and  we set  $L=100$. The inset shows the  order parameters over  a  
larger range,  confirming   the  analytical  result   $V=1$  in the   strong-coupling limit [Eq.~\eqref{MF_strong_coupling.eq}].   (b)  Free energy  as function of~$\Jk$.  Both (a) and  (b)  support  a  first-order  transition.}
\label{fig:Mft_phase_diagram}
\end{figure}

%The resulting mean-field phase diagram is presented in Fig.~\ref{fig:Mft_phase_diagram}. Here, we have set the hopping parameter for the  conduction electrons to  $t=1$ to define the unit energy, fixed the Heisenberg coupling  to  $\Jh/t=0.5$, and varied the Kondo  coupling  $\Jk$.    The phase diagram is divided into three regimes.  At weak Kondo coupling, $\Jk\ll 1$, we observe an antiferromagnetic metallic phase,  in which the mean-field parameters satisfy $m_{c}\ne 0$, $m_{f}\ne 0$, and $V=0$.  At strong Kondo coupling, $\Jk \gg 1$, the model is in a    paramagnetic  heavy-fermion  phase,   characterized by $m_{c} = m_{f}=0$ and $V\ne0$. These two phases are separated  by a coexistence  region  centered  around  $\Jk\approx3.45$, in which all three mean-field parameters are finite, $m_{c} \ne 0$, $m_{f} \ne 0$, and $V \ne0$.

The resulting mean-field phase diagram at zero temperature is presented in Fig.~\ref{fig:Mft_phase_diagram}(a). Here, we have set the hopping parameter for the  conduction electrons to  $t=1$, thereby setting the unit of energy, fixed the Heisenberg coupling  to  $\Jh/t=0.5$, and varied the Kondo  coupling  $\Jk$. The phase diagram is divided into two regimes.  At weak Kondo coupling, $\Jk\ll 1$, we observe an antiferromagnetic metallic phase,  in which the mean-field parameters satisfy $m_{c}\ne 0$, $m_{f}\ne 0$, and $V=0$. At strong Kondo coupling, $\Jk \gg 1$, the model is in a    paramagnetic  heavy-fermion  phase,   characterized by $m_{c} = m_{f}=0$ and $V\ne0$.  These two phases are separated  by a direct first-order transition around  $\Jk\approx3.42$, where the order parameters  show  discontinuities. A  cusp in the  ground state, Fig.~\ref{fig:Mft_phase_diagram}(b) reflects the corresponding level crossing,  consistent with the first-order nature of the transition.

\begin{figure}[tb]
\begin{centering}
\includegraphics[width=\linewidth]{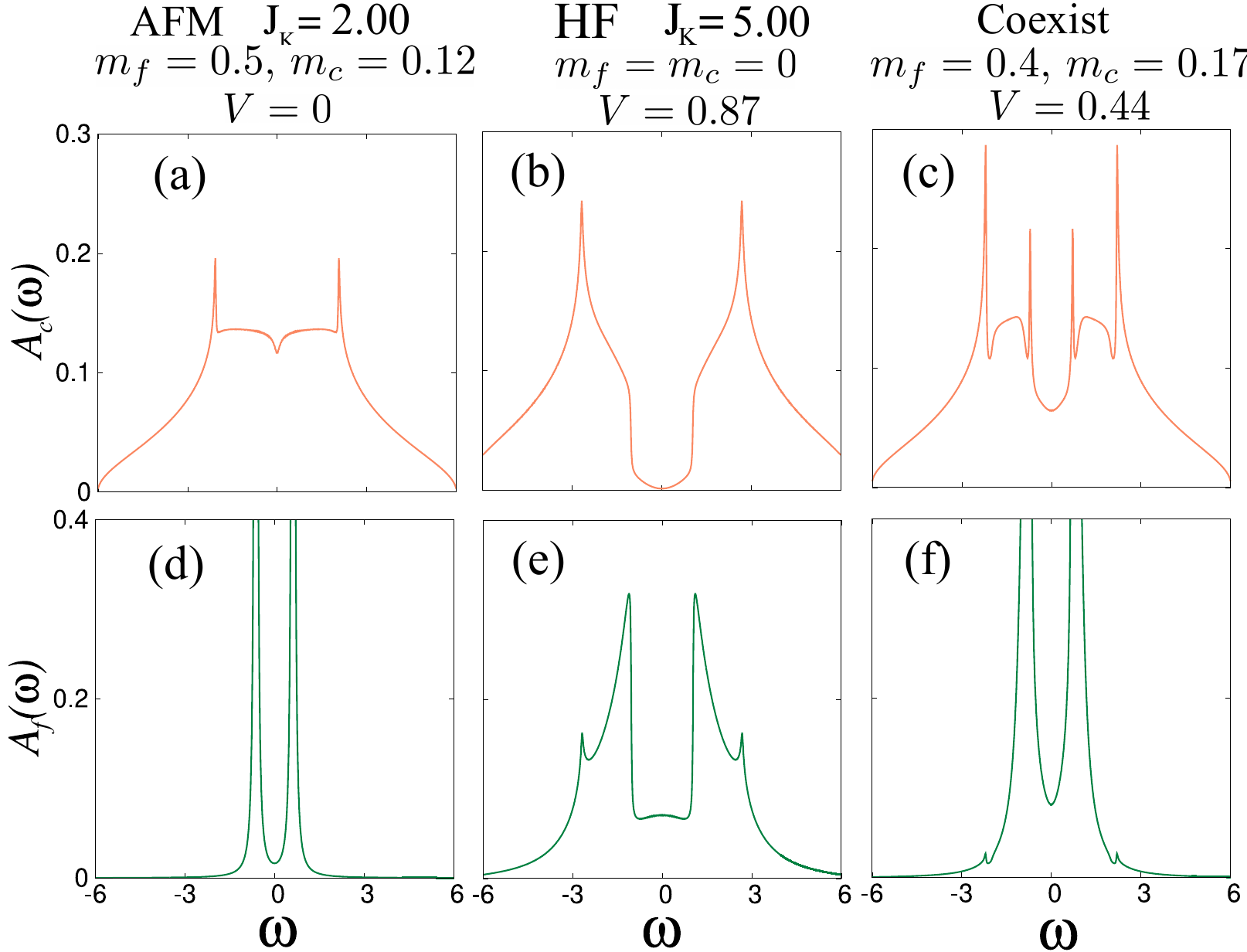}
\par\end{centering}
\caption{Local density of states $A_{c}(\omega)$  (top) and $A_{f}(\omega)$ (bottom) in antiferromagnetic  (first column),  heavy-fermion (second column) and coexistence (third column) states in mean-field approximation. The corresponding mean-field parameters are given in the title of each column.}
\label{fig:Mft_dos}
\end{figure}

We expect   that  the single-particle spectral   function  $A_{c/f} (\ve{k}_2,\omega) =- \frac{1}{\pi}\text{Im}  G^{\text{ret}}_{c/f} (\ve{k}_2,\omega )$ will have distinct features  in each phase.  Here,  $G_{d}^{\text{ret}}  (\ve{k}_2, \omega) = - i \int_{0}^{\infty} dt \, e^{i \omega t}  \sum_{\sigma } \left<\big\{\hat{d}_{\ve{k}_2,\sigma}^{}(t), \hat{d}_{\ve{k}_2,\sigma}^{\dagger} (0) \big\} \right>$, with $\hat{d}_{\ve{k}_2,\sigma} = \hat{c}_{\ve{k}_2,R_z=0,\sigma} $  for the  conduction-electron spectral  function, and $\hat{d}_{\ve{k}_2,\sigma}  =  \hat{f}_{\ve{k}_2,\sigma}$ for the pseudo-fermion spectral function. In Fig.~\ref{fig:Mft_dos},   we plot  the local  density  of  states, $A_{c/f}(\omega)= \frac{1}{L^2} \sum_{\ve{k}_2} A_{c/f}(\ve{k}_2,\omega)$  in the  aforementioned  two phases, using $\Jk=2.00$ and $\Jk = 5.00$. Furthermore, we are interested in states characterized by all order parameters being non-zero. Although such state are not realized as ground states at the level of the mean-field approximation, it reflects the coexistence of magnetic order and  Kondo 
screening,  as observed in the   forthcoming quantum Monte Carlo results.
%In the rest of this section, we define the coexistence region, characterized by non-zero values of $m_c$, $m_f$, and $V$, and present the results of the single-particle spectral function in the mean field level as a reference.}
 
\begin{figure}[tb]
\begin{centering}
--\includegraphics[width=\linewidth]{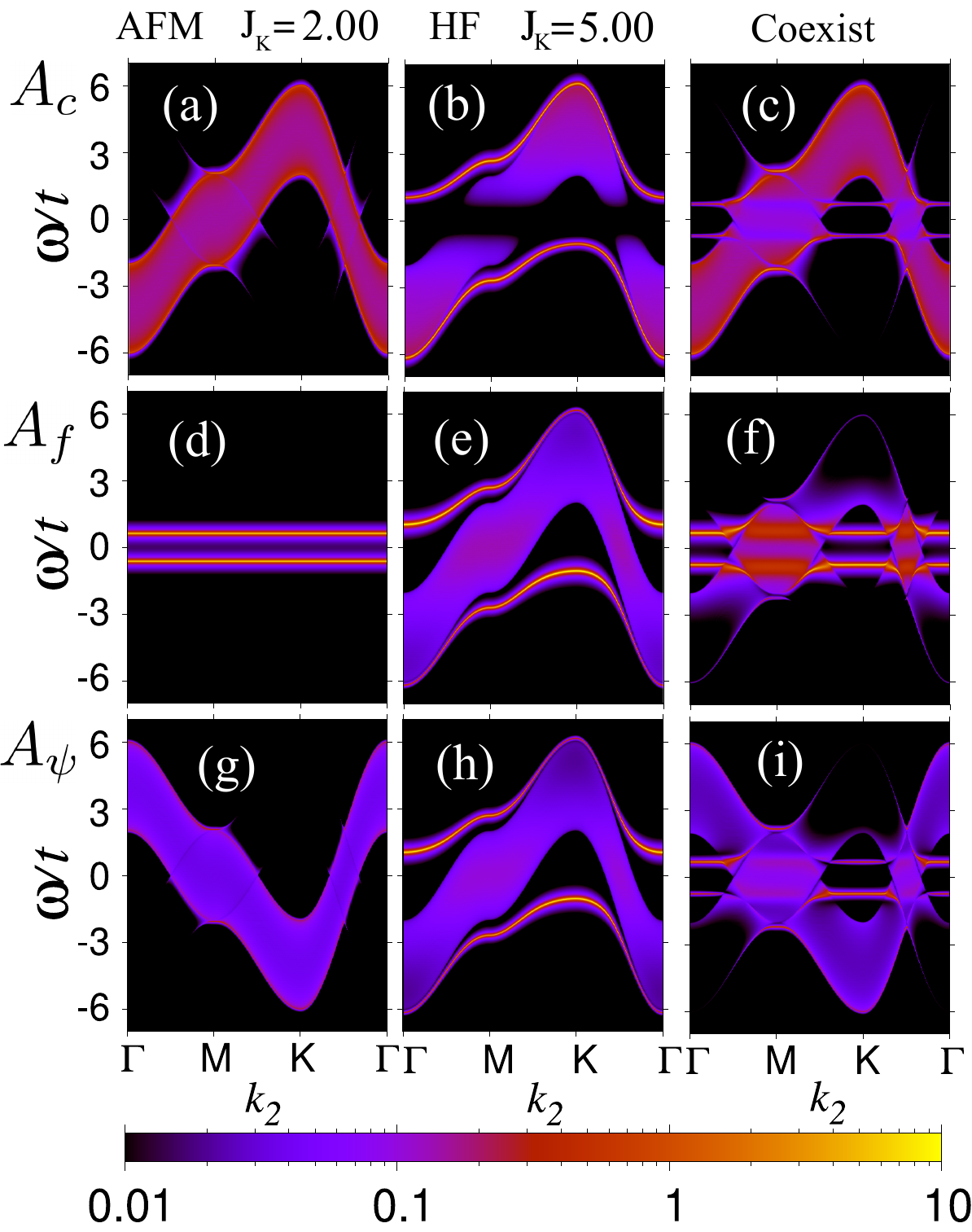}
\par\end{centering}
\caption{Fermion spectral functions $A_{c}(\ve{k}_2,\omega)$ (top row), $A_{f}(\ve{k}_2,\omega)$ (center row), and $A_{\psi}(\ve{k}_2,\omega)$ (bottom row) in antiferromagnetic (left column), heavy-fermion (center column) and coexistence (right column) phases in mean-field approximation. Here, we  consider the path  $\ve{\Gamma}(0,0)\rightarrow \ve{M}(\pi,0)\rightarrow \ve{K}(\pi,\pi)\rightarrow\ve{\Gamma}(0,0)$.}
\label{fig:Mft_akw}
\end{figure}

For $m_c\ne 0$, $m_f  \neq 0$, and $V = 0$, as  observed  at    $\Jk=2.00$,   the $f$ fermions  are  localized.  $A_{f}(\omega)$, shown in Fig.~\ref{fig:Mft_dos}(d),   consists  of  two  Dirac $\delta$ functions  and  the origin of  the  gap  stems from the Weiss  mean field $m_f \neq 0$. This is  confirmed  by the  momentum-resolved  $f$ spectral function, shown in Fig.~\ref{fig:Mft_akw}(d), which
exhibits  two   flat  bands.
For  the  above mean-field parameters,  the  conduction-electron   resolvant  matrix   reads
$ G_{c,\sigma}(z)   =   \left( G_{c,\sigma,0}^{-1}(z) -  \Sigma_{c,\sigma} \right)^{-1}$
with  
$\left[ G_{c,\sigma,0}^{-1}(z)   \right]_{\ve{k},\ve{k}' } =  \delta_{\ve{k},\ve{k}'}  (z  - \epsilon(\ve{k})) $, $\left[ \Sigma_{c,\sigma} \right]_{\ve{k},\ve{k}'}   
 =  \frac{\sigma \Jk m_f}{2L}   \delta_{\ve{k}_2,\ve{k}'_2   + \ve{Q} } $,
 and  $z$  a  complex  frequency.   Here,   $ \ve{k}   =  \left( \ve{k}_2, k_z \right)   $  and   $\ve{Q} =  (\pi,\pi) $. From this  form,  one  can  derive the  spectral  function,    plotted in  Fig.~\ref{fig:Mft_akw}(a),
 \begin{equation}\label{eq:AcMF}
 	A_c(\ve{k}_2,\omega)      = -  \frac{1}{\pi}  \text{Im}  \frac{1}{g_0^{-1}(\ve{k}_2,\omega) -   \frac{(\Jk m_f)^{2}}{4} \, g_0( \ve{k}_2+ \ve{Q},  \omega) } 
 \end{equation}
 with   $ g_0( \ve{k}_2,\omega) =    \frac{1}{L}   \sum_{k_z}   \frac{1}{ \omega  +  i 0^{+}     -  \epsilon(\ve{k}_2,k_z)} $.     In the  absence  of  magnetic  ordering,  the  spectral  function  describes   a continuum of  extended   Bloch  states  in  all  three  directions.    At  finite  values of  $m_c$,  we observe     a  back  folding of this  structure  due  to  scattering  off  the magnetic Bragg peak.  In the vicinity  of  the  $\ve{M}$ point and   $\omega \simeq \pm 2.1t$,    
  we  observe a  pole  that  is  detached  from the  continuum.  Note  that  at the $\ve{M}$ point, the  continuum of states  obtained 
  from  $g_0( \ve{k}_2,\omega)$    is  bounded  by  $\omega =  \pm 2$.    Let  $|\ve{k}_2,  k_z \rangle $  be  the  wave  function  corresponding  to  the pole.  Since   the  two-dimensional  $\ve{k}_2$ vector  is  a  conserved  quantity, up  to a  reciprocal lattice  vector of the magnetic Brillouin zone, an  electron in this state cannot  decay  into  an  extended  three-dimensional  state.   Hence, we  expect  the  wave function to  have  a two-dimensional character.  That is,  $ | \ve{k}_2, R_z \rangle \equiv   \frac{1}{\sqrt{L}}  \sum_{k_z} e^{- i k_z R_z} | \ve{k}_2,  k_z \rangle $  should decay  exponentially  as  a function of  $R_z$, with $R_z=0$  denoting  the magnetic  layer.      In  Appendix \ref{mean_field.app},  we provide an  explicit  calculation in the paramagnetic phase,  demonstrating this point.  
  The two dimensionality  of   the  pole  shows  up in  Fig.~\ref{fig:Mft_dos}(a).  
Here, we see   the  saddle point of the dispersion at  the  $\ve{M}$ wave vector leads  to  a  two-dimensional van-Hove  singularity  with  characteristic logarithmic  divergence.\footnote{In one dimension, a van-Hove  singularity would lead to a  square-root  singularity,  while in a three-dimensional translational-invariant system, it would lead to a  cusp  or  singularity in the first  derivative.}

 At  large   $\Jk  =  5.00$, shown in Figs.~\ref{fig:Mft_dos}(b,e)  and  \ref{fig:Mft_akw}(b,e), only   $V$  takes a  nonvanishing  value.  Here  the  $f$ electron  delocalizes  and  participates in the Luttinger  volume. This notion  can  be  made  precise,    since  we  can  view the  heterostructure  as  a  two-dimensional Bravais  lattice  with a unit cell  consisting  of  
 $L$ conduction electrons   and  one  pseudo-fermion.
In the paramagnetic heavy-fermion phase, the pseudo-fermion spectral  function   is given   by
 \begin{equation}\label{eq:A_fMF}
 	A_f(\ve{k}_2,\omega)      = -  \frac{1}{\pi}  \text{Im}   \frac{1}{ \omega + i0^+  -  \left(\Jk V/2\right)^2 g_{0}(\ve{k}_2,\omega) }. 
 \end{equation}
For  values of  $\ve{k}_2$   with  $\cos k_x  + \cos k_y = 0$,  defining   the Fermi  surface of the two-dimensional  tight-binding  model on the square lattice,   the  above form  matches   the  mean-field  result  of  a single impurity   in a one-dimensional  metallic  host.    Here,  we  observe  a  resonance at 
the Fermi  energy   for the $f$ spectral   function, and  a dip in the $c$ spectral function. These  features  are  apparent  in   Figs.~\ref{fig:Mft_dos}(b,e) and \ref{fig:Mft_akw}(b,e).    The momentum-resolved  $f$ spectral function,  \ref{fig:Mft_akw}(e), shows  well-defined poles.  Following the same  discussion as  above, and  as  explicitly  computed  in Appendix \ref{mean_field.app},  these poles  correspond  to  two-dimensional  states.    In the vicinity of the $\ve{M}$ and  $\ve{\Gamma}$ points in the two-dimensional Brillouin zone,   they   form  a narrow  band  that  leads to an enhanced  density  of  states,  very  visible  in  Fig.~\ref{fig:Mft_dos}(b,e).

Finally, for a coexistence state, all  mean-field   order  parameters  take  nonvanishing  values. For our analysis, we choose the mean-field parameters $m_c=0.17$, $m_f=0.40$, and $V=0.438$. The  dominant features  in such state, as shown in  Figs.~\ref{fig:Mft_dos}(c,f) and \ref{fig:Mft_akw}(c,f),   can be  understood  by  starting  from the  paramagnetic heavy-fermion phase and    allowing  for  $\ve{Q} =  (\pi,\pi) $  scattering,  which  leads  to  shadow  bands in the  extended zone  scheme.   
 
In  an exact  numerical  calculation, we  do not have access to the   pseudo fermion,   and  it is convenient to  consider  a so-called  composite fermion 
operator defined as $\hat{\psi}_{\ve{i},\sigma}^{\dagger}=\sum_{\sigma^{\prime}}\hat{c}_{\ve{i},R_{z}=0,\sigma^{\prime}}^{\dagger}\boldsymbol{\sigma}_{\sigma^{\prime},\sigma}^{} \cdot\hat{\boldsymbol{\mathit{S}}}^f_{\ve{i}}$ \cite{Costi00,Borda07,Maltseva09,Raczkowski18,Danu21,Raczkowski22}.   
By  means of  a  canonical  Shrieffer-Wolff transformation \cite{Schrieffer66}, one  can  derive the Kondo lattice  model   from  an  Anderson model  in the limit where charge fluctuations on the localized impurity  orbitals  is  suppressed. 
  In this framework, the  composite fermion operator merely  corresponds to the Schrieffer-Wolff transformation of   the  fermion  creation operator
  on localized   impurity orbitals \cite{Raczkowski18}.
In addition, one can represent the Kondo coupling as the hybridization of the composite fermion and $c$ fermion,  $\frac{1}{2}\sum_{\sigma} ( \hat{\psi}_{\ve{i},\sigma}^{\dagger}\hat{c}_{\ve{i},R_z=0,\sigma}^{}+\hat{c}_{\ve{i},R_z=0,\sigma}^{\dagger}\hat{\psi}_{\ve{i},\sigma}^{}) =\hat{\boldsymbol{S}}_{\ve{i},R_z=0}^{c}\cdot\hat{\boldsymbol{S}}_{\ve{i}}^{f}$. 
In the mean-field approximation, the Green's function of the composite fermion can be computed by expressing it as a convolution of single-particle Green's functions via Wick's theorem. The resulting momentum-dependent spectral function $A_{\psi}(\ve{k}_2,\omega)$ is depicted for the three representative values of $\Jk$ in Figs.~\ref{fig:Mft_akw}(g)-(i).

We first focus on the behavior of the composite fermion in a Kondo-screened phase. In the large-$N$ limit, the composite fermion reads $\hat{\psi}^{\dagger}_{\ve{i},\sigma}\propto \left\langle\hat{D}^{}_{\ve{i}}\right\rangle\frac{2}{N} f^{\dagger}_{\ve{i},\sigma}$\cite{Danu21}, where $N$ is the number  of fermion components, 
$\sigma = 1,\dots,N$,
and $\hat{D}_{\ve{i}}=\sum_{\sigma=1}^{N}\hat{f}_{\ve{i},\sigma}^{\phantom\dagger} \hat{c}_{\ve{i},R_{z}=0,\sigma}^{\dagger}$  the hybridization. The Kondo-screened phase is characterized by a finite hybridization parameter $V$,  and we expect the spectral function of $\psi$ fermion to follow that of the $f$ fermion. By comparing the results presented in Figs.~\ref{fig:Mft_akw}(e) and (h), we see that the mean-field calculation agrees with the large-$N$ approximation in the Kondo-screened phase.

In the antiferromagnetic metal phase, the mean-field hybridization parameter $V$ vanishes, so that  the $f$ electron is no longer related to the composite fermion $\psi$. The behavior of the composite fermion  spectral function  in Fig.~\ref{fig:Mft_akw}(g) can be understood within a  large-$S$ approximation~\cite{Danu21}. 
 Using the Holstein-Primakoff representation of the spin algebra and  at  lowest order in  $1/S$, we obtain $\sum_{\sigma}G^{\psi}_{\ve{i},\ve{j}}(\tau)=S^2\sum_{\sigma}\left\langle\hat{c}_{\ve{i},\sigma}^{}(0)\hat{c}_{\ve{j},\sigma}^\dagger(\tau)\right\rangle e^{i\boldsymbol{Q}\cdot(\ve{j}-\ve{i})}$. Hence, the composite fermion Green's function can be obtained from  that of the $c$ fermions,  albeit with momentum shifted by  the  magnetic  wave vector $\boldsymbol{Q}$. This statement is verified upon comparing Figs.~\ref{fig:Mft_akw}(a) and \ref{fig:Mft_akw}(g).

%%%%%%%%%%%%%%%%%%%%%%%%%%%%%%%%%%%%%%%%%%%%%%%%%%%%%%%%%%%%%%%%%%%%%%%

\section{QMC simulations}\label{sec:QMC}
We use the ALF~\cite{ALF_v2,Assaad08_rev} implementation of  the    finite-temperature  \cite{Blankenbecler81,Hirsch85,White89}   and projective  \cite{Sugiyama86,Sorella89}   auxiliary-field  QMC  algorithms to perform large-scale simulations of the  model defined in Eq.~\eqref{eq:model}.    For  the QMC  simulations,   we  consider  the  Hamiltonian,   
\begin{align} 
\hat{H}_{\text{QMC}} & = -t\sum_{\ve{k}_2,\sigma,R_{z},R_{z}^{\prime}}\hat{c}_{\ve{k}_2,R_{z},\sigma}^{\dagger}T(\ve{k}_2)_{R_{z},R_{z}^{\prime}}^{\phantom\dagger} \hat{c}_{\ve{k}_2,R_{z}^{\prime},\sigma}^{\phantom\dagger}
\nonumber \allowdisplaybreaks[1]\\& \quad 
- \frac{\Jk}{4}\sum_{\ve{i}}\left(\sum_{\sigma}\hat{c}_{\ve{i},R_{z}=0,\sigma}^{\dagger}\hat{f}_{\ve{i},\sigma}^{}+\hat{f}_{\ve{i},\sigma}^{\dagger}\hat{c}_{\ve{i},R_{z}=0,\sigma}^{}\right)^{2}
\nonumber\allowdisplaybreaks[1]\\& \quad 
- \frac{\Jh}{4}\sum_{\left\langle \ve{i}, \ve{j} \right\rangle }\left(\sum_{\sigma}\hat{f}_{\ve{i},\sigma}^{\dagger}\hat{f}_{\ve{j},\sigma}^{}+\hat{f}_{\ve{j},\sigma}^{\dagger}\hat{f}_{\ve{i},\sigma}^{}\right)^{2}
\nonumber\allowdisplaybreaks[1]\\& \quad 
+\frac{U}{2}\sum_{\ve{i}}\left(\sum_{\sigma}\hat{f}_{\ve{i},\sigma}^{\dagger}\hat{f}_{\ve{i},\sigma}^{}-1\right)^{2},
\label{eq:model_qmc}
\end{align}
with $T(\ve{k}_2)_{R_z, R_z'}=-2t(\cos k_{x} + \cos k_{y}) \delta_{R_z,R_z^{\prime}}-t\delta_{|R_z-R_z^{\prime}|,1}-t(\delta_{R_z,L-1}\delta_{R^{\prime}_z,0}+\delta_{R_z,0}\delta_{R^{\prime}_z,L-1})$
and $\hat{f}_{\ve{i},\sigma}$   the  pseudo-fermion operator.  The Hubbard-$U$ interaction in Eq.~\eqref{eq:model_qmc} suppresses charge fluctuations of the pseudo fermion.  Crucially,   the  local $f$-fermion parity,  $(-1)^{ \hat{n}^f_{\ve{i}} }$,  is a  conserved  quantity,  such  that  the  unphysical  even-parity  states  are  suppressed  exponentially  as  $\beta U$   grows. 
In the odd-parity  sector,  $\hat{H}_\text{QMC}$ is equivalent to the KLM-hetero Hamiltonian, $\left.\hat{H}_{\text{QMC}}\right|_{\hat{n}_{\ve{i}}^{f} = 1}=\hat{H}_{\text{KLM-hetero}}$. 
In the practical finite-temperature (projective) simulations,  
we keep the product $\beta U > 10$ ($2\Theta U>10$, where $\Theta$ is the projection length), which is sufficient to suppress the charge fluctuations of the ${f}$ fermion.

Eq.~\eqref{eq:model_qmc} provides a U(1)-gauge-theory description of the Kondo-lattice problem. Following the path-integral formalism used in Ref.~\cite{Raczkowski22},  we introduce
bosonic fields $b_{\ve{i}}(\tau)\propto \boldsymbol{c}^{\dagger}_{\ve{i}}(\tau)\boldsymbol{f}_{\ve{i}}(\tau)$ and $b^f_{\ve{i},\ve{j}}
(\tau)\propto\boldsymbol{f}^{\dagger}_{\ve{i}}(\tau)\boldsymbol{f}_{\ve{j}}(\tau)$ by decomposing the perfect square terms parametrized by $\Jk$ and $\Jh$ in Eq.~\eqref{eq:model_qmc},
where $\boldsymbol{c}_{\ve{i}}(\tau)$ and $\boldsymbol{f}_{\ve{i}}(\tau)$ are the Grassmann fields of the $c$ and $f$ fermion operators.  At $U=\infty$, the action has local U(1) gauge invariance. $\boldsymbol{f}_{\ve{i}}(\tau)$ and $b_{\ve{i}}(\tau)$ are the U(1)-gauge-charged field variables. One can define the gauge-neutral composite fermion field as $\boldsymbol{\tilde{f}}_{\ve{i}}(\tau)=e^{i\varphi_{\ve{i}}(\tau)}\boldsymbol{f}_{\ve{i}}(\tau)$ with $e^{i\varphi_{\ve{i}}(\tau)}=\frac{b_{\ve{i}}(\tau)}{\left|b_{\ve{i}}(\tau)\right|}$. The composite fermion field carries electron charge and spin $1/2$. More importantly, the field variable $\boldsymbol{\tilde{f}}_{\ve{i}}(\tau)$ corresponds to the Grassmann variable of the composite fermion operator $\boldsymbol{\tilde{f}}_{\ve{i}}\propto\boldsymbol{\psi}_{\ve{i}}=\boldsymbol{S}_{\ve{i}}\cdot\boldsymbol{\sigma}\boldsymbol{c}_{\ve{i}}$. This relation provides a route to understand the dynamic-spin-correlation behavior from the heavy-fermion band structure. We will come back to this point in the next section. 

In the QMC simulations, we treat the finite-size Kondo heterostructure as a two-dimensional lattice containing $L\times L$ unit cells with $L+1$ orbitals per in-plane unit cell. The orbitals refer to the $z$-axis degrees of freedom. We use two techniques to reduce the finite-size effects in the lattice simulations. One is to follow the technique suggested in Ref.~\cite{Assaad01}, by including an orbital  magnetic  field  of  strength  $B  = \phi_0/L^2$  in the  $z$ direction.  Another technique we use is to   average  over  twisted boundary conditions in the $z$ direction \cite{Gross92}.  To  achieve  this, we  consider  a  distinct twist on every  process during the parallel computations.  Averaging  over parallel  runs  then  amounts  to averaging  over all possible twisted boundary conditions.  The details of this  technique are  presented in Appendix~\ref{twist.app}.

%%%%%%%%%%%%%%%%%%%%%%%%%%%%%%%%%%%%%%%%%%%%%%%%%%%%%%%%%%%%%%%%%%%%%%%

\section{Results}\label{sec:QMC_result}

In this section, we  discuss  our QMC results.  We  consider  three-dimensional   systems  with  linear lattice sizes  $L=4,6,8,10,12$,  and  a mix  of finite-temperature and  zero-temperature  projective methods.  In our simulations, we set $t=1$   to  define  the unit  energy,   fix $\Jh/t = 0.5$    and vary  the  Kondo  coupling. For projective  QMC simulations,  we   consider  the  projection length parameter $\Theta=40$ to ensure  convergence of the results.

\subsection{Phase diagram}

\begin{figure}[tb]
\begin{centering}
\includegraphics[width=\linewidth]{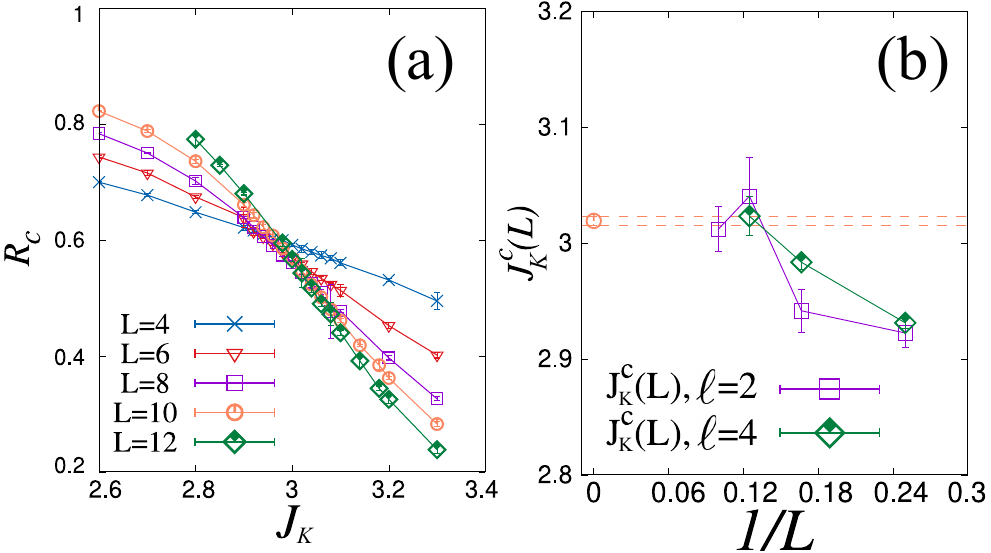}
\par\end{centering}
\caption{(a) Correlation ratio $R_\mathrm{c}(L,\Jk)$ as function of Kondo coupling $\Jk$ from projective QMC for different lattice sizes $L$. (b) Finite-size critical point $\Jkc(L)$ as function $1/L$. The extrapolation towards the thermodynamic limit indicates a single quantum critical point at $\Jkc=3.019(4)$.}
\label{fig:rc_t0}
\end{figure}

From the  mean-field analysis, we anticipate  at least one magnetic   quantum phase  transition as a  function of  $\Jk$.   To pin down the location of a possible phase transition,  we  use  a  renormalization-group-(RG)-invariant quantity, the  correlation ratio,  given  by  
\begin{equation}
R_\mathrm{c}\left( \left(\Jk -  \Jkc\right) L^{1/\nu}, L^{z}/\beta, L^{-\omega} \right)=1-\frac{C_{f}(\ve{Q}+d\ve{k}_2,0)}{C_{f}(\ve{Q},0)}
\end{equation}
where $C_{f}(\ve{k}_2,0)$ is the spin structure factor of  the  impurity  spins,  
\begin{equation} 
C_{f}\left(\ve{k}_2,\tau\right)=\sum_{\ve{k}_2}e^{-i \ve{k}_2 \cdot(\ve{i} -  \ve{j})} \left\langle \hat{S}_{\ve{i}}^{f}(\tau)\hat{S}_{\ve{j}}^{f}(0)\right\rangle,
\end{equation}
and $d\ve{k}_2=\left(2\pi/L,0\right)$ is the smallest momentum difference on the  finite-size system. The RG-invariant quantity $R_\mathrm{c}$ approaches  one  in the magnetically-ordered phase and drops to zero   for  short-ranged spin  correlations.    Since   we  \emph{a priori}  do  not know the value of the dynamical  exponent $z$,   we  have used the projective   algorithm,  such  that  we  can set $L^z/\beta = 0 $  in the above equation.  If the corrections to scaling are  small (i.e., $\omega$ is large), we  expect  a  universal  crossing  at $\Jkc$.   In Fig.~\ref{fig:rc_t0}(a),  we plot  the  result of $R_\mathrm{c}$ for  $L=4,6,8,10,12$ at zero temperature. 
The finite-size critical points $\Jkc(L)$ is defined  by the intersection of the  correlation ratio between  different system sizes,  
$R_\mathrm{c}(\Jkc(L), L)=R_\mathrm{c}(\Jkc(L),L+\ell)$.  In Fig.~\ref{fig:rc_t0}(b),  we consider  $\ell=2,4$, and  use a polynomial  fit  to    determine the crossing  point. 
 We  see  that   $\Jkc(L)$  drifts as a function of  growing size and   stabilizes  to 
$\Jkc=3.019(4)$ at the two largest system sizes in our calculation.
As we will demonstrate below, in contrast to the mean-field result, the QMC spectral functions indicate that hybridization between conduction electrons and local moments occurs throughout the  magnetic  phase.
The QMC phase diagram therefore consists of just two different phases: An antiferromagnetic heavy-fermion phase for small $\Jk < \Jkc$ and a paramagnetic heavy-fermion phase for large $\Jk > \Jkc$, see Fig.~\ref{fig:plot_of_model}(d).

\subsection{Spin spectral function}

\begin{figure}[tb]
\begin{centering}
\includegraphics[width=\linewidth]{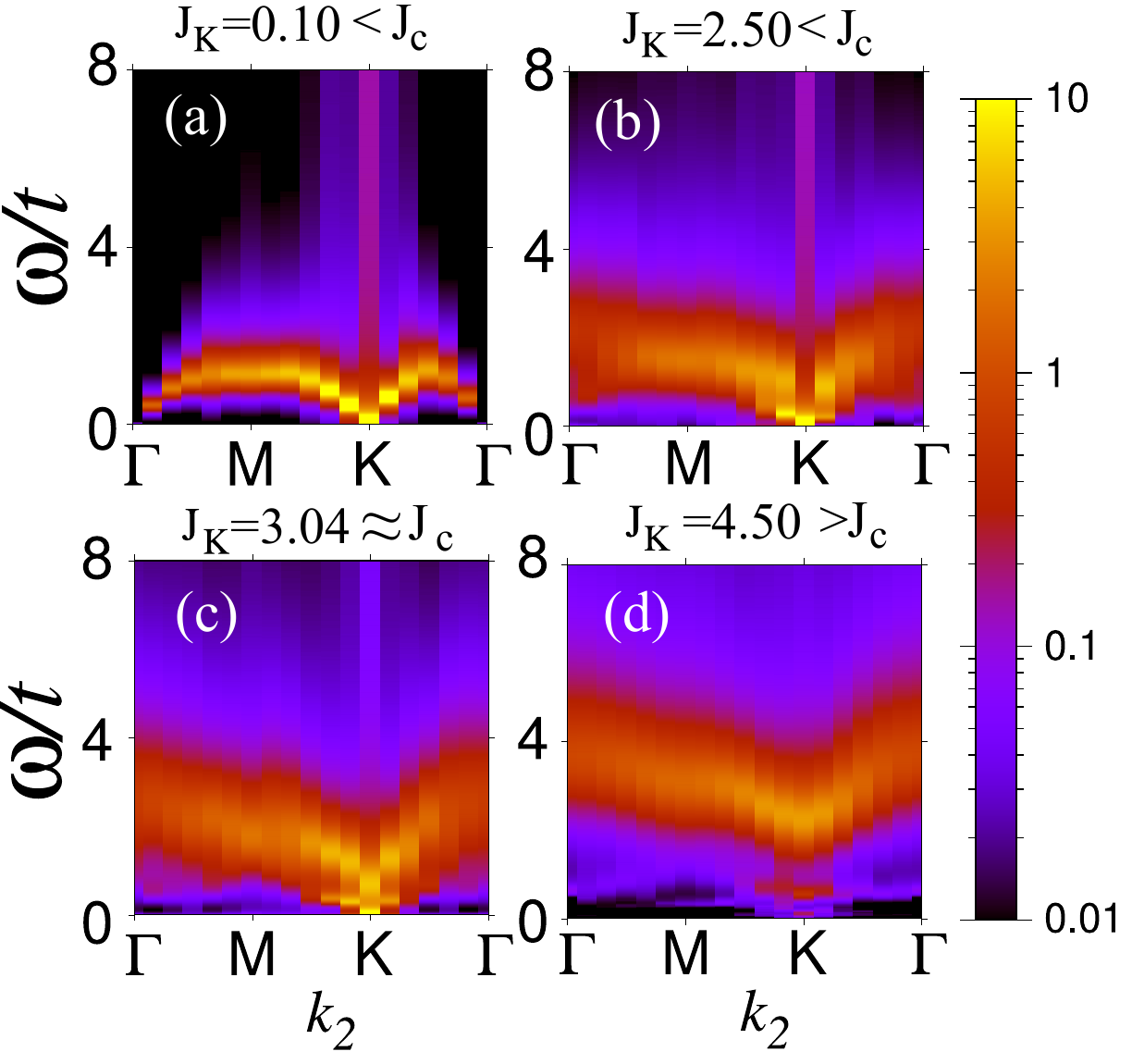}
\par\end{centering}
\caption{Spin spectral function $S(\ve{k}_2, \omega)$ along a high-symmetry path in the Brillouin zone from finite-temperature QMC for different values of $\Jk$, using $\beta=40$ and $L=12$.}
\label{fig:Spin_akw}
\end{figure}

We turn our   attention  on the spin susceptibility  of  the  magnetic impurity layer.  Using the ALF \cite{ALF_v2} implementation of the stochastic  Maximum Entropy method \cite{Sandvik98,Beach04a}, we extract the the dynamical spin structure factor $S(\ve{k}_2,\omega)=\chi''_f(\ve{k}_2,\omega)/[1-\exp(-\beta\omega)]$ from imaginary time spin correlation function.  Specifically, 
\begin{equation}
	C_{f}\left(\ve{k}_2,\tau\right) = \frac{1}{\pi} 
   \int d \omega \, \frac{e^{- \tau \omega} }{ 1 - e^{-\beta  \omega} } \chi''_f(\ve{k}_2,\omega).
\end{equation}
 At weak coupling, such as  $\Jk=0.50$, Fig.~\ref{fig:Spin_akw}(a),  the  dominant features  of the   dynamical
  spin structure factor follow  the  spin-wave result  with a linear mode around the ordering wave vector $\ve{Q}=(\pi,\pi)$.  Upon  increasing  
  the  Kondo  coupling to  $\Jk=2.50$, Fig.~\ref{fig:Spin_akw}(b), which is still in the magnetically-ordered  phase,  we  observe  
  marked   differences  from the  spin-wave result:   
  The spectral weight broadens  and  the  data is consistent  with    a  lower edge of the  spectra  that  follows 
$\omega \propto (\ve{k}_2 - \ve{Q})^2$ near $\ve{Q}$.   This is  consistent with  the  notion of Landau-damped  Goldstone  modes  originating   from the Kondo coupling of the spins  to the metallic  host, as presented in  Sec.~\ref{Weak_coupling:sec}  and  discussed  in  Appendix \ref{spin_wave.app}.  Fig.~\ref{fig:Spin_akw}(c) demonstrates that this 
  feature is  apparent up to the critical point $\Jkc$.   
 At strong coupling, $\Jk=4.50 > \Jkc$, Fig.~\ref{fig:Spin_akw}(d),  we  are in the paramagnetic  heavy-fermion  phase.  The absence of long-range magnetic  order  with associated  Bragg peaks at  the  antiferromagnetic  wave  vector  leads  to a  strong suppression of low-energy weight.   We, however,  still  observe  low-energy  spectral  weight:     The  heavy-fermion  phases  are  characterized by  the emergence  of  the composite  fermion operator.  As  mentioned  above, and  within  a  field-theoretical framework,  this  can be understood in terms  of a Higgs mechanism  in which   $e^{i\varphi_{\ve{i}}(\tau)} f_{\ve{i}}(\tau) $   
  is a  well-defined  low-energy  excitation.  The low-energy   spectral  weight  corresponds  to the  particle-hole bubble of this   composite fermion. 
   On the other   hand,  the high-energy  spectral  weight  is  reminiscent of the    Kondo  insulator \cite{Capponi00,Tsunetsugu97_rev}  that  captures  triplon   dynamics.   

\begin{figure}[tb]
\begin{centering}
\includegraphics[width=0.96\linewidth]{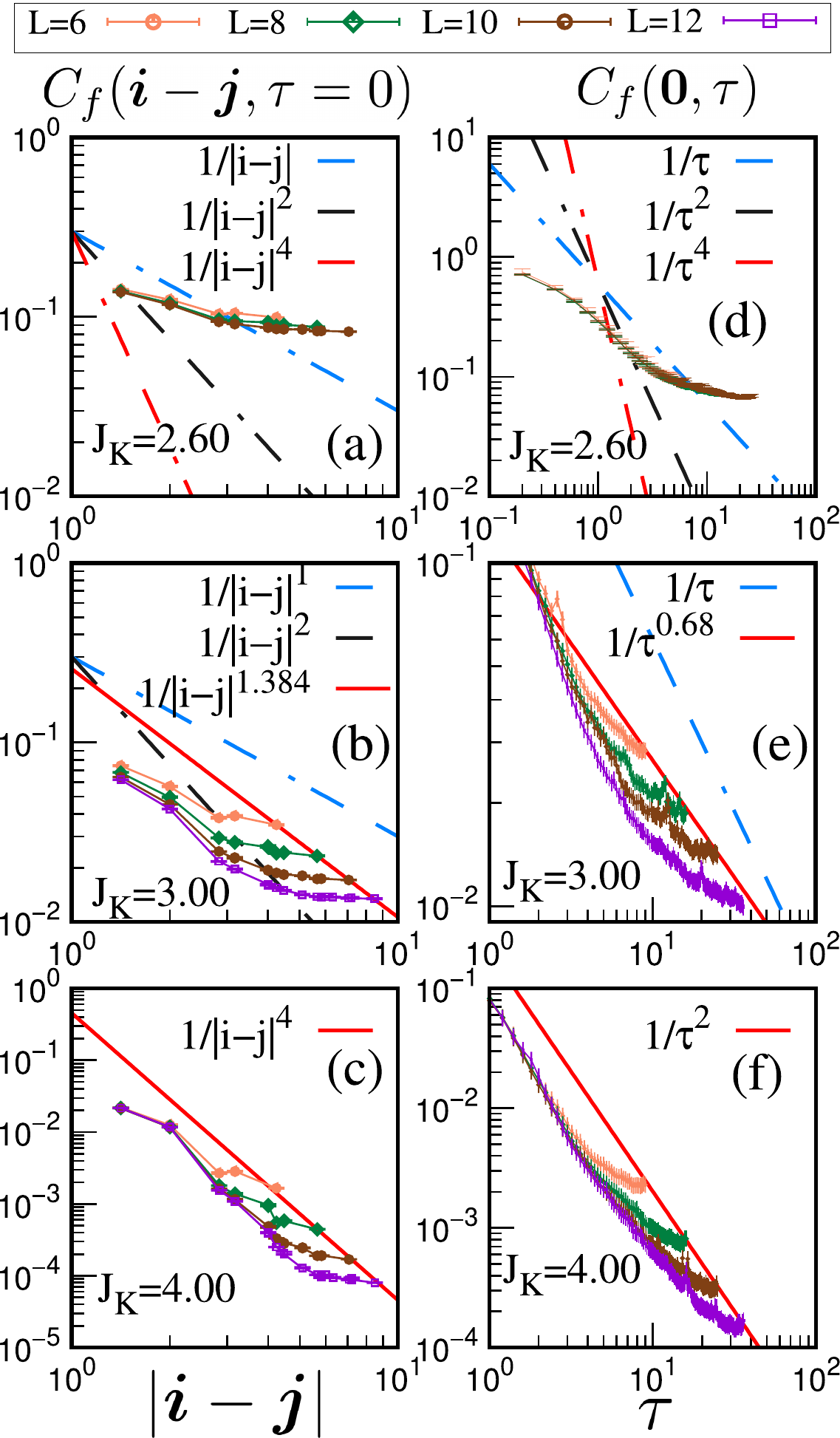}
\par\end{centering}
\caption{Spin-spin correlations $C_f(\ve{i}-\ve{j},\tau)$ in impurity layer as function of distance $r$ at equal time $\tau = 0$ (left column) and imaginary time $\tau$ at equal position $r = 0$ (right column) for different values of $\Jk$ in the antiferromagnetic phase (top row), quantum critical regime (center row), and paramagnetic heavy-fermion phase (bottom row), from finite-temperature QMC, using $\beta=L^2/2$.  Dashed blue, black, and red lines represent power-law decay functions $f(x)\sim 1/x$, $1/x^2$ and $1/x^4$ for reference. The red solid line at (b)(c)(e)(f) represent the numerical fitting of the QMC data.}
\label{fig:chirtau}
\end{figure}

Following the weak-coupling-limit discussion and the observation of Landau damping  in  the dynamical spin structure factor,  we foresee that the dynamical exponent  is given by $z=2$ at the magnetic critical point $\Jkc$.    Since  the  dynamical  exponent encodes  the  asymmetry   between  space and 
time,  we  consider 
real-space equal-time spin correlations $C_f(\ve{i}-\ve{j},\tau=0)$ in Figs.~\ref{fig:chirtau}(a)-(c), as  well  as  local time-displaced correlations $C_f(\ve{0},\tau)$ in Figs.~\ref{fig:chirtau}(d)-(f).   Here,  
\begin{equation}
	C_f(\ve{i}-\ve{j},\tau)  =  \left\langle \hat{S}_{\ve{i}}^{f}(\tau)\hat{S}_{\ve{j}}^{f}(0)\right\rangle.
\end{equation}
For  this set  of  calculations,  we  have used  the  finite-temperature  code  with $\beta=L^2/2$ so as  to  observe  ground-state  properties.  
 At $\Jk=2.60<\Jkc$,  Figs.~\ref{fig:chirtau}(a,d), the Heisenberg model  has  long-range order, such that   the  spin  correlations  saturate  to  a  constant  both in space  and  
 imaginary time.      We note that   our  simulations  explicitly preserve  SU(2)  spin symmetry,  so  that we  cannot  distinguish  between  longitudinal and  transverse modes.  In other words,   the  data of  Figs.~\ref{fig:chirtau}(a,d)    are   dominated     by  the  long-range   order,  and  we are  blind  to the 
 transverse critical modes. 

 At the critical point $\Jk=3.00\approx \Jkc$, Figs.~\ref{fig:chirtau}(b,e),    the  data suggest power-law decays of the form $C_f(\ve{i}-\ve{j},\tau=0)\propto 1/|\ve{i}-\ve{j}|^{\alpha_r}$  and  $C_f(\ve{0},\tau)\propto 1/\tau^{\alpha_t}$, respectively.  By performing a numerical fit, we have extracted the values of the exponents, resulting in $\alpha_r=1.384(2)$ and $\alpha_t=0.68(2)$, satisfying $\alpha_r=2\alpha_t$ within numerical uncertainty, consistent with $z=2$ at the critical point. The fits are represented as solid red lines in Figs.~\ref{fig:chirtau}(b,e).  

Finally, in the paramagnetic heavy-fermion phase, Figs.~\ref{fig:chirtau}(c,f),   the  spin-spin  correlations  inherit  the  scaling of the host  metal.  That is,  $C_f(\ve{i}-\ve{j},\tau=0)\propto 1/|\ve{i}-\ve{j}| ^4$ in  space  and  $C_f(\ve{0},\tau)\propto 1/\tau^2$   in imaginary  time.   We  understand  this   from the point of  view of the 
composite fermion operator,%
\footnote{Since  $\ve{\tilde{f}}_{i}(\tau)$  is only  defined  within the  path integral,  $\ve{\tilde{f}}_{i}(\tau)$ in the second line of Eq.~\eqref{eq:composite-fermion-correlation} corresponds  to  a  Grassmann variable and  we consider  $\tau > 0$. }
\begin{align} \label{eq:composite-fermion-correlation}
\left\langle \hat{\boldsymbol{S}}_{\ve{i}}^{f}(\tau) \hat{\boldsymbol{S}}_{\ve{j}}^{f}(0)\right\rangle  & =\left\langle \frac{1}{2}\hat{\boldsymbol{f}}_{\ve{i}}^{\dagger}(\tau)\boldsymbol{\sigma}\hat{\boldsymbol{f}}_{\ve{i}}(\tau) \cdot\frac{1}{2}\hat{\boldsymbol{f}}_{\ve{j}}^{\dagger}(0)\boldsymbol{\sigma}\hat{\boldsymbol{f}}_{\ve{j}}(0) \right\rangle \nonumber \allowdisplaybreaks[1] \\ 
& =\left\langle \frac{1}{2}\boldsymbol{\tilde{f}}_{\ve{i}}^{\dagger}(\tau)\boldsymbol{\sigma}\boldsymbol{\tilde{f}}_{\ve{i}}(\tau)\cdot\frac{1}{2}\boldsymbol{\tilde{f}}_{\ve{j}}^{\dagger}(0)\boldsymbol{\sigma}\boldsymbol{\tilde{f}}_{\ve{j}}(0)\right\rangle.
\end{align}
In the  paramagnetic  heavy-fermion  phase,  the   spin  correlations are  well   understood  by  considering the bubble of the  above particle-hole    correlation  function.  In fact, in the large-$N$ limit, vertex  contributions  vanish,   and  as  shown in Ref.~\cite{Raczkowski20} for the   specific  case of  the half-filled   two-dimensional Kondo lattice model,  the large-$N$  saddle point is adiabatically  connected to the SU(2) model.  Since  the  $\ve{\tilde{f}}_{\ve{i}}(\tau) $  operator  has  the  same  quantum numbers  as  that of the electron operator,  we expect it to  have the same scaling  dimension.

\subsection{Composite-fermion and  conduction-electron spectral functions}
\begin{figure*}[t!]
\includegraphics[width=\linewidth]{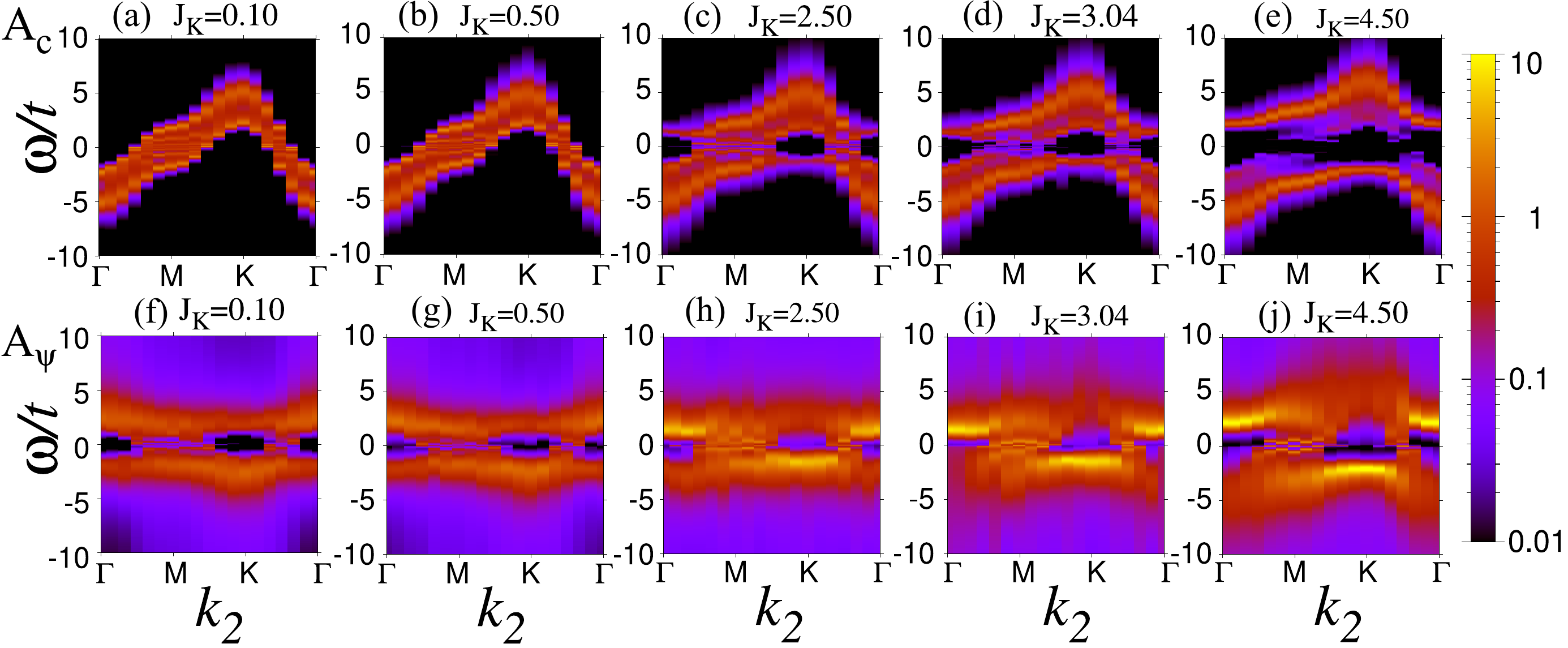}
\caption{Fermion spectral functions $A_c(\ve{k}_2,\omega)$ (top row) and $A_{\psi}(\ve{k}_2,\omega)$ (bottom row) from finite-temperature QMC, using $\beta=40$ and $L=12$. The Kondo coupling $\Jk$ increases from left to right.}
\label{fig:cot_grc_plot}
\end{figure*}

The  composite-fermion  spectral  function  is a  very  useful  quantity  to  assess  the  presence  of  Kondo screening.   Let us  start  with the  corresponding  periodic Anderson model.     In this case,  Kondo  breakdown  corresponds  to an orbital-selective Mott  transition \cite{Vojta10}, and  the
single-particle spectral  function of the   impurity-orbital fermion operator $\hat{d}^{\dagger}$  will  show a gap.   In the limit  where  charge  fluctuations on the  impurity orbitals are  suppressed  and the  periodic Anderson model maps onto the Kondo lattice model,    the  composite   fermion is  nothing  but the  canonical  
Schrieffer-Wolff  transformation of the   $\hat{d}^{\dagger}$  operator. Hence, Kondo  breakdown corresponds  to an absence  of spectral  weight  at the  Fermi  energy of the composite  fermion operator. 

In Fig.~\ref{fig:cot_grc_plot}, we present the evolution of the  $c$-fermion spectral function $A_{c}(\ve{k}_2,\omega)$ and the composite-fermion spectral function $A_{\psi}(\ve{k}_2,\omega)$ upon varying the  Kondo coupling $\Jk$.  At  weak coupling, $\Jk=0.1$ and $\Jk = 0.5$, the magnetic impurities   exhibit long-range antiferromagnetic  order. The $c$-fermion spectral function  is  very  similar  to the corresponding mean-field result.  The composite-fermion spectral function $A_{\psi}(\ve{k}_2,\omega)$ reveals a momentum shift  of $\bold{Q}=(\pi,\pi)$  with respect to $A_c(\ve{k}_2,\omega)$, see also Fig.~\ref{fig:Mft_akw}(g). In addition, the intense composite-fermion spectral weight at $\omega \simeq 0$ at the $\ve{\Gamma}$ point suggests Kondo screening throughout the antiferromagnetic phase for all $\Jk > 0$. This feature becomes clear by comparing Figs.~\ref{fig:cot_grc_plot}(f,g) with the mean-field composite-fermion spectral function in the coexistence phase, Fig.~\ref{fig:Mft_akw}(i).

As one   enhances  the Kondo coupling into the finite-temperature quantum critical fan, $\Jk=2.50$ and $\Jk = 3.04$, we observe  growing (decreasing) low-energy  spectral  weight  in the  composite-fermion ($c$-electron)  spectral function. 
 Due to the  reduction of the antiferromagnetic order parameter,  band folding features in the composite-fermion spectral  function 
 become weaker as compared to smaller values of $\Jk$.

\begin{figure}[htb]
\begin{centering}
\includegraphics[width=\linewidth]{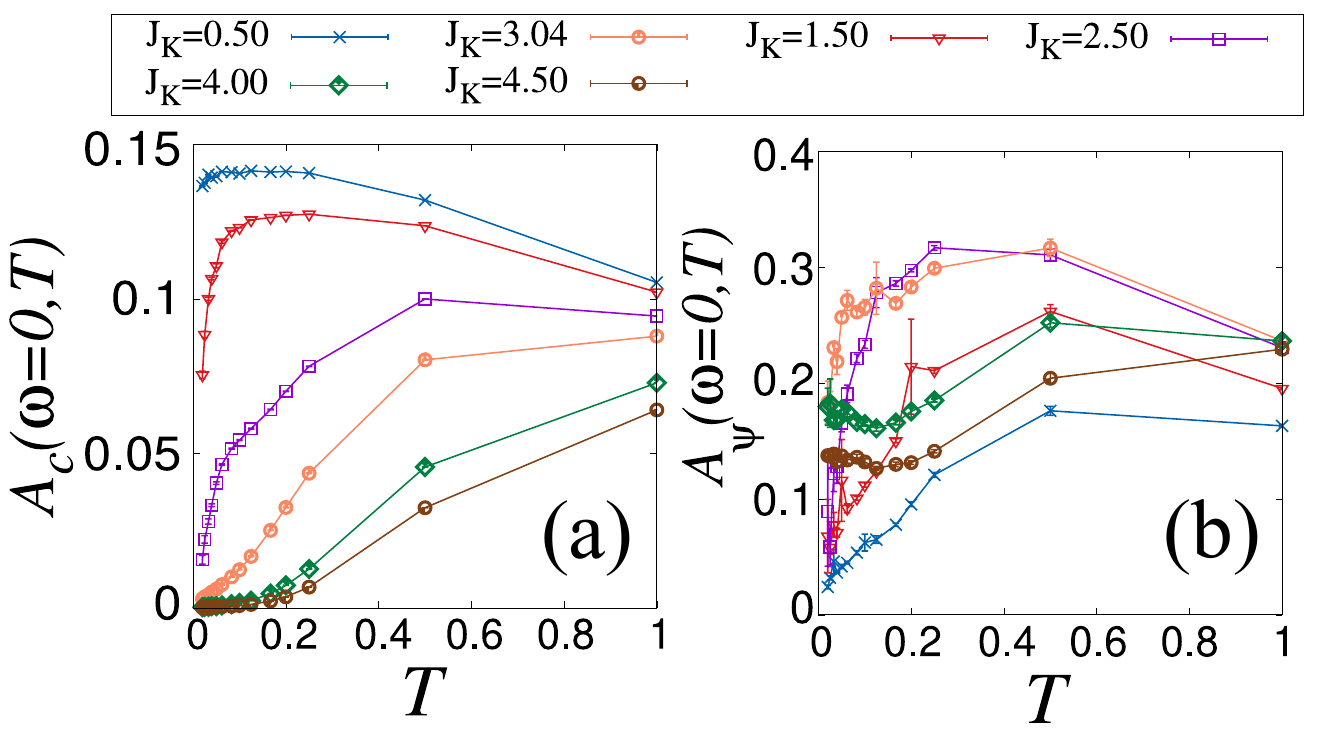}
\par\end{centering}
\caption{Local density of states $A_c(\omega=0)$ (left) and $A_\psi(\omega=0)$ (right) near the Fermi level as function of temperature $T = 1/\beta$ for  different values of $\Jk$ from finite-temperature QMC.   Here,  we  consider $L=10$ and restrict the data to the temperature range with smallest finite-size effects.}
\label{fig:dos_cf_T}
\end{figure}

At large Kondo coupling, $\Jk=4.50$,  deep in the paramagnetic  heavy-fermion phase,  magnetic correlations are  short  ranged  and  the  $c$ fermion strongly hybridizes with the $f$ pseudo fermion.  In Fig.~\ref{fig:cot_grc_plot}(e), we observe that the  low-energy $c$-fermion spectral weight is  greatly  suppressed.   This is  consistent  with the mean-field  result  of Fig.~\ref{fig:Mft_akw}(b). 
On the  other hand,  Fig.~\ref{fig:cot_grc_plot}(j)   shows  that  the composite fermion has substantial  low-energy  weight, again in accordance  to the 
large-$N$  calculation of  Fig.~\ref{fig:Mft_akw}(h).

As  mentioned at the beginning of this section, it is  crucial to  understand  whether the $f$-fermion spectral  function  has  finite   spectral weight at the Fermi  energy, since  this a  measure  for  Kondo screening.   In Fig.~\ref{fig:dos_cf_T}, we present the temperature dependence of the local density states near the Fermi level $A_{c/{\psi}}(\omega=0,T)$ 
as function of temperature. Here we use the approximate relation $A_{c/{\psi}}(\omega=0)\approx\frac{\beta}{\pi N}\sum_{k_2}G_{c/{\psi}}\left(k_2,\tau=\frac{\beta}{2}\right)$ to obtain this quantity directly, circumventing the need for analytical continuation.  Note that for a  smooth density of  states  at the Fermi  level,   this equation  becomes exact in the low-temperature limit.  
At low temperatures, as presented in Fig.~\ref{fig:dos_cf_T}(a), the $c$-fermion local density of states, $A_{c}(\omega=0)$, decreases  with 
growing  Kondo coupling. When $\Jk>\Jkc$,  $A_{c}(\omega=0)$  becomes  very small at  low  temperatures. In contrast,  the results for the composite-fermion local density of states $A_{\psi}(\omega=0)$, as shown in Fig.~\ref{fig:dos_cf_T}(b), suggest that this quantity does not vanish for  any finite $\Jk$ in the zero-temperature limit.
Since  composite and $c$  fermions have  the same quantum  numbers,  the  supports  of both spectral  functions  are  expected to  be identical.  However,  the spectral  weight can  differ  substantially.   Hence,  we  understand  that the  low-energy  spectral  weight of  the conduction electrons   deep in the paramagnetic  heavy-fermion phase does  not vanish. 
Further data, demonstrating that these results are representative of the thermodynamic limit, are provided in Appendix~\ref{finite_size.app}.

On the  whole,  the    results shown  in this section  provide numerical evidence of  a metal-to-metal    magnetic  transition   across  which Kondo screening does \emph{not}  break down.

%%%%%%%%%%%%%%%%%%%%%%%%%%%%%%%%%%%%%%%%%%%%%%%%%%%%%%%%%%%%%%%%%%%%%%%

\section{Conclusions and outlook}\label{sec:Conclusion}

The model of the  Kondo heterostructure  presented in this work  provides a unique possibility to numerically investigate   the physics  of  quantum spins
in a  metallic  environment  without  encountering the infamous negative-sign problem.  The model  can be  seen as  a  dimensional generalization 
of  a  spin-chain on a metallic  surface \cite{Danu20,Danu22},  leading to a  two-dimensional quantum antiferromagnet embedded in  a 
 three-dimensional  metal.  Our  model  is relevant for  the  description of  Kondo  heterostructures such as  CeIn$_3$/LaIn$_3$ superlattices studied   experimentally  in  Ref.~\cite{shishido10}. 

Combining a weak-coupling analysis  and a mean-field calculation, we foresee  the  existence of a magnetic  quantum critical point  in a  
metallic  environment driven by  the  Kondo  interaction.   This is confirmed  by unbiased large-scale  auxiliary-field QMC simulations. 
The  antiferromagnetic  heavy-fermion phase  is   characterized  by  Landau-damped   Goldstone  modes  and  the  quantum  critical  point is consistent  with a   dynamical  exponent  $z=2$.  Both  aspects  are  a  direct  consequence of  the metallic  environment.    In the paramagnetic heavy-fermion phase,  the spin correlations  of  the  magnetic  system inherit 
those   of  the host  metal, in accordance  with  the  large-$N$  calculation.   This result  can be understood  in terms of  the emergence of a composite  fermion operator  that  carries  the   quantum number of the electron and hence  possesses  the  same  scaling   dimension.    Within  a  U(1)  gauge  theory of the Kondo lattice,    the  composite  fermion  corresponds to the  bound state of the  Abrikosov  pseudo fermion
and the phase  of  the bosonic  hybridization field. 

Of  crucial  importance  for the understanding of the transition is the fate  of the aforementioned  composite fermion and  the associated   Kondo effect. 
In fact, up to the smallest Kondo coupling we considered, $\Jk=0.1$, the composite-fermion  spectral  function  does  not  develop a  gap,   such  that 
we  can  exclude   Kondo  breakdown.   We  note  that Kondo  breakdown within  the magnetically-ordered phase would not   violate Luttinger's theorem  due  to the  doubling of the magnetic unit cell \cite{vojta08}. 
Hence, the quantum critical point in our model describes an interesting metal-to-metal magnetic transition in a model with SU(2) local spins,  in which 
the  heavy-fermion quasiparticle does  neither  disintegrate at the transition nor in the magnetic phase.  Consequently,  this  transition  falls into  the  category  
of  Hertz-Millis \cite{Hertz76,Millis93},  albeit  with the important property  that   only  the two-dimensional  crystal momentum  is  conserved  up to a reciprocal lattice vector. The  understanding of  this transition and  a possible  non-Fermi liquid  character  is  left  for  future  work.   Another  intriguing  issue is  the  finite-temperature  phase  diagram.    In the  very  same  way  that  dissipation   stabilizes   long-range  order  in the ground state of a  one-dimensional  spin-chain \cite{Weber22}, one might  expect  the  Kondo  heterostructure  to  show   magnetism at  finite  temperature.  

%%%%%%%%%%%%%%%%%%%%%%%%%%%%%%%%%%%%%%%%%%%%%%%%%%%%%%%%%%%%%%%%%%%%%%%

\begin{acknowledgments}
FA    and  MV   acknowledge  enlightening conversations  with  T.  Grover   and B.  Danu  on   related  subjects. The authors gratefully acknowledge the Gauss Centre for Supercomputing e.V.\ (www.gauss-centre.eu) for funding this project by providing computing time on the GCS Supercomputer SUPERMUC-NG at Leibniz Supercomputing Centre (www.lrz.de).
This research has been supported by the Deutsche Forschungsgemeinschaft through the W\"urzburg-Dresden Cluster of Excellence on Complexity and Topology in Quantum Matter -- \textit{ct.qmat} (EXC 2147, Project No.\ 390858490), SFB 1143 on Correlated Magnetism (Project No.\ 247310070), SFB 1170 on Topological and Correlated Electronics at Surfaces and Interfaces (Project No.\  258499086), and the Emmy Noether Program (JA2306/4-1, Project No.\ 411750675).
\end{acknowledgments}

\appendix

%%%%%%%%%%%%%%%%%%%%%%%%%%%%%%%%%%%%%%%%%%%%%%%%%%%%%%%%%%%%%%%%%%%%%%%
\section{Numerical solution of self-consistency equations}
\label{self_consistency.app}

In the mean-field analysis, we employ the standard iterative method to solve the self-consistency equations, Eq.~\eqref{eq:mft_selfconsist}. We denote the set of mean-field parameters at the $n$-th step of the iteration as $x_n$, with $n \in \mathbb N$.
The iterative method starts with an initial guess of the mean-field order parameters $x_1=( m_{c,1}, m_{f,1}, V_1 )$. At each step of the iteration, we compute the single-particle Green's function by diagonalising the mean-field Hamiltonian $\hat H_\text{MF}$, Eq.~\eqref{eq:mft_ham}, with input parameters $x_n$. The convergence is determined using a quantity $\rho^2_n$, defined as
\begin{equation}
\rho^2_n=\rho_{c,n}^{2}+\rho^{2}_{f,n}+\rho^{2}_{V,n},
\end{equation}
where
\begin{align}
\rho_{c,n} & =m_{c,n} + \frac{1}{2L^{2}} \sum_{\ve{k}_{2},\sigma}\left(-1\right)^{\sigma} 
\nonumber \\ &\quad
\times \left\langle \hat{c}_{\ve{k}_{2}+\ve{Q},R_{z}=0,\sigma}^{\dagger}\hat{c}_{\ve{k}_{2},R_{z}=0,\sigma}^{\phantom{\dagger}}\right\rangle_n, \\
\rho_{f,n} & =m_{f,n}-\frac{1}{2L^{2}}\sum_{\ve{k}_{2},\sigma}\left(-1\right)^{\sigma}\left\langle \hat{f}_{\ve{k}_{2}+\ve{Q},\sigma}^{\dagger}\hat{f}_{\ve{k}_{2},\sigma}^{\phantom{\dagger}}\right\rangle_n,\\
\rho_{V,n} & =V_n-\frac{1}{2L^{2}}\sum_{\ve{k}_{2},\sigma}\left\langle \hat{c}_{\ve{k}_{2},R_{z}=0,\sigma}^{\dagger}\hat{f}_{\ve{k}_{2},\sigma}^{\phantom{\dagger}}+\text{h.c.}\right\rangle_n, 
\end{align}
where $\langle \dots \rangle_n$ denotes expectation value with respect to $\hat H_\text{MF}$ at the $n$-th step of the iteration, i.e., with input parameters $x_n$.
If $\rho^2_n$ is larger than or equal to a small threshold value $r \ll 1$, the algorithm proceeds to the next iteration, and we update the mean-field parameters $x_{n+1}$ according to Eq.~\eqref{eq:mft_selfconsist}, using the Green's function obtained in the previous step. For $\rho^2_n < r$, we have obtained a fixed-point solution $x_n$ with a precision of $r$. In our analysis, we set the threshold value of $r$ to $10^{-10}$, which is sufficient for a system size of $L=100$.

%%%%%%%%%%%%%%%%%%%%%%%%%%%%%%%%%%%%%%%%%%%%%%%%%%%%%%%%%%%%%%%%%%%%%%%

%%%%%%%%%%%%%%%%%%%%%%%%%%%%%%%%%%%%%%%%%%%%%%%%%%%%%%%%%%%%%%%%%%%%%%%

\section{Twisted boundary conditions in $z$ direction}
\label{twist.app}

For periodic boundary conditions, the fermion operator satisfies $\hat{c}_{\ve{r}+\ve{L}_{a}}=\hat{c}_{\ve{r}}$,
where $a=1,2,3$ represents  the  $x,y,z$ directions. 
For twisted boundary conditions along the $z$ direction, the condition transforms
into $\hat{c}_{ \ve{r}+\ve{L} _{a}}=e^{ 2 \pi i \Phi/\Phi_0  \delta_{a,z}} \hat{c}_{\ve{r} }$.  The  twist in the  boundary  can  be removed  at  the  expense  of  a   vector  potential  in the  Hamiltonian  that   can be locally  but  not  globally   removed  with a  gauge  transformation \cite{Byers61}.        Specifically  we  can consider  the  canonical  transformation: 
\begin{equation}
	   \tilde{d}_{\ve{r}}   =     e^{ -   \frac{2 \pi i  \Phi}{\Phi_0}  \frac{\ve{e}_z \cdot \ve{r}}{L} }  \tilde{c}_{\ve{r}}. 
\end{equation}
$ \tilde{d}_{\ve{r}}  $   satisfies  periodic  boundary  conditions, 
\begin{equation}
	  \tilde{d}_{\ve{r}  + \ve{L}_a}    =     \tilde{d}_{\ve{r} }
\end{equation}
and  the  hopping  part  of  our  Hamiltonian   transforms  to: 
\begin{equation}
	    \hat{H}_{\text{Fermi}}  =    - t  \sum_{\langle  \ve{r}, \ve{r}'  \rangle }    \hat{d}^{\dagger}_{\ve{r}}  e^{ \frac{2\pi i  \Phi}{\Phi_0}  \frac{ \left(  \ve{r}-\ve{r}' \right) \cdot  \ve{e}_z}{L}}  \hat{d}^{\phantom\dagger}_{\ve{r}'} 
	    \label{eq:H0_k_twist}
\end{equation}

Fourier  transform gives: 
\begin{equation}
	\hat{H}_{\text{Fermi}} =  \sum_{\ve{k}}\left[-2t \sum_{a=1}^{3} \cos\left(k_{a} + \delta_{a,3} \frac{\Phi}{\Phi_{0}}\frac{2\pi }{L}\right)\right]\hat{d}_{\ve{k}}^{\dagger}\hat{d}^{}_{\ve{k}}.
\end{equation}

In the QMC calculation, we obtain the observable by
averaging over different twisted boundary condition $\Phi/\Phi_{0}\in\left[0,1\right]$
to reduce finite-size effects.    In particular  for  a  one-dimensional  non-interacting system   it  
was  pointed out in Ref.~\cite{Gross92} such an  averaging  over  boundary  conditions  yields  exact 
results  for any  value of $L$.  

\begin{figure}[tb]
\begin{centering}
\includegraphics[width=\linewidth]{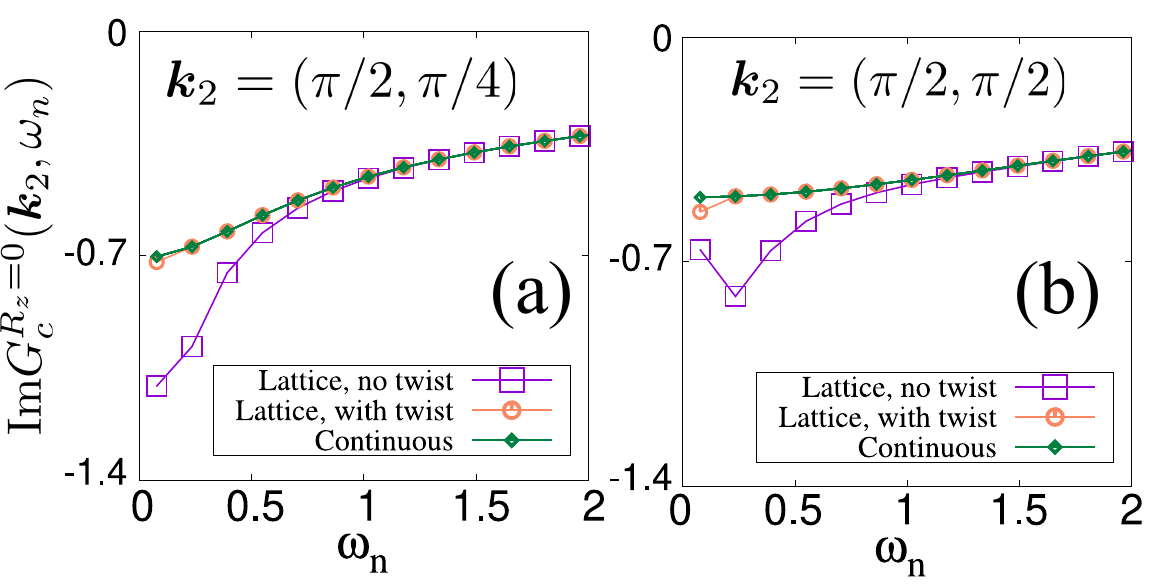}
\par\end{centering}
\caption{Imaginary part of $c$-fermion Green's function $\Im G_c^{R_z = 0}(\boldsymbol{k}_2,\omega_n)$ as function of Matsubara frequency $\omega_n=(2n+1)\pi/\beta$ for (a) $\ve k_2 = (\pi/2,\pi/4)$ and (b) $\ve k_2 = (\pi/2,\pi/2)$. Green dots follow Eq.~(\ref{eq:gloc0}). Purple (orange) dots are obtained from finite-size lattice calculations with linear size $L=8$ and inverse temperature $\beta=40$ with periodic boundary conditions (twisted boundary conditions, averaged over ten different twists). Lines represent guides to the eye.}
\label{fig:grc0_compare}
\end{figure}

Hence,  for a given operator
$\hat{O}$,  we  evaluate:  
\begin{equation}
\left\langle \hat{O}\right\rangle =\int_{0}^{1}d\Phi \left\langle \hat{O}\left(\Phi \right)\right\rangle. 
\end{equation}
This strategy improve the data quality in the free fermion system by increasing the momentum resolution for finite system sizes. In Fig.~\ref{fig:grc0_compare}, we provide a simple benchmark of the noninteracting $c$-fermion Green's function at the impurity layer, obtained from $\hat{H}_{\text{Fermi}}$, defined as 
\begin{equation}
G^{R_z = 0}_{c,0}(\ve{k}_2,\omega_n)=-\int_0^{\beta} d\tau\,e^{i\omega_n\tau}\left\langle \hat{c}_{\ve{k}_2,R_{z}=0}(\tau) \hat{c}_{\ve{k}_2,R_{z}=0}^{\dagger}  (0)\right\rangle_0
\end{equation}
where $\omega_n=(2n+1)\pi/\beta$ is the Matsubara frequency. By considering Eq.~(\ref{eq:H0_k_twist}) in continuous momentum space, $G^{R_z = 0}_{c,0}(\ve{k}_2,\omega_n)$ follows the analytic form given in Eq.~(\ref{eq:gloc0}), see the green line in Fig.~\ref{fig:grc0_compare}. In the lattice calculation, this quantity suffers from the finite momentum resolution and deviates from the analytic form at low frequency, which is well observed for periodic boundary conditions, shown in purple in Fig.\ref{fig:grc0_compare}.
The orange dots shown in Fig.~\ref{fig:grc0_compare} represent the results of finite-size lattice calculations with twisted boundary conditions. In this calculation, we average over ten different twists, satisfying $\Phi/\Phi_0=0.1n$ for $n\in[0,1,\dots,9]$. As presented in Fig.\ref{fig:grc0_compare}, the mismatch between the finite-size calculation and the analytic form of $G^{R_z = 0}_{c}(\ve{k}_2,\omega_n)$ at low frequency can be reduced effectively by using twisted boundary conditions, at least in the noninteracting system. On this ground, we believe this technique can alleviate  finite-size  effects also in the interacting system.

%%%%%%%%%%%%%%%%%%%%%%%%%%%%%%%%%%%%%%%%%%%%%%%%%%%%%%%%%%%%%%%%%%%%%%%

%%%%%%%%%%%%%%%%%%%%%%%%%%%%%%%%%%%%%%%%%%%%%%%%%%%%%%%%%%%%%%%%%%%%%%%
\section{Finite-size effects on local density of states}
\label{finite_size.app}

\begin{figure}[tb]
\begin{centering}
\includegraphics[width=\linewidth]{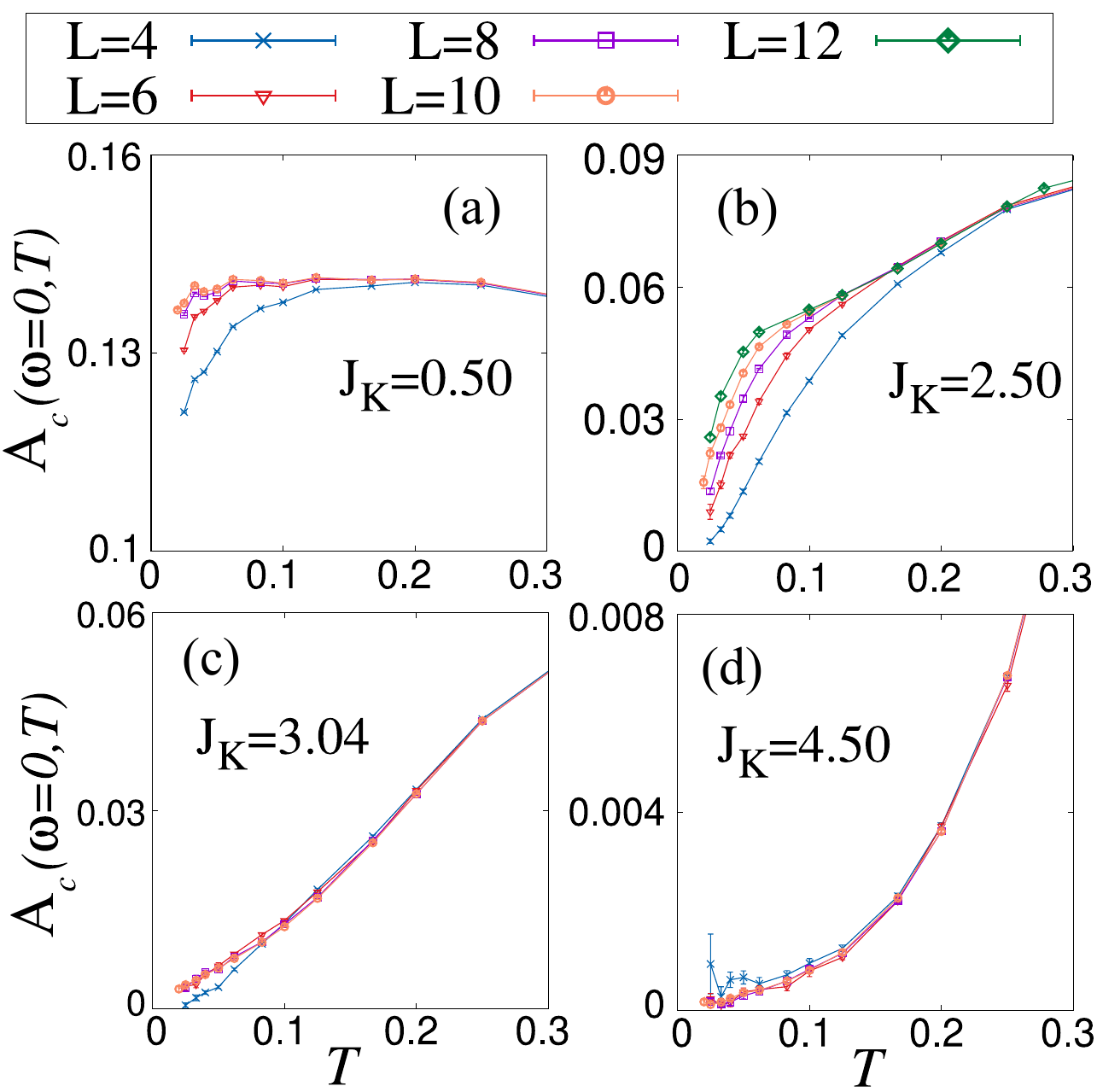}
\par\end{centering}
\caption{Conduction-electron local density of states $A_c(\omega=0,T)$ near the Fermi level as function of temperature $T$ for different lattice sizes $L$. In each panel, different colors indicate different system sizes. The Kondo coupling $\Jk$ increases from left to right and top to bottom.}
\label{fig:cdos_size_effect}
\end{figure}

\begin{figure}[tb]
\begin{centering}
\includegraphics[width=\linewidth]{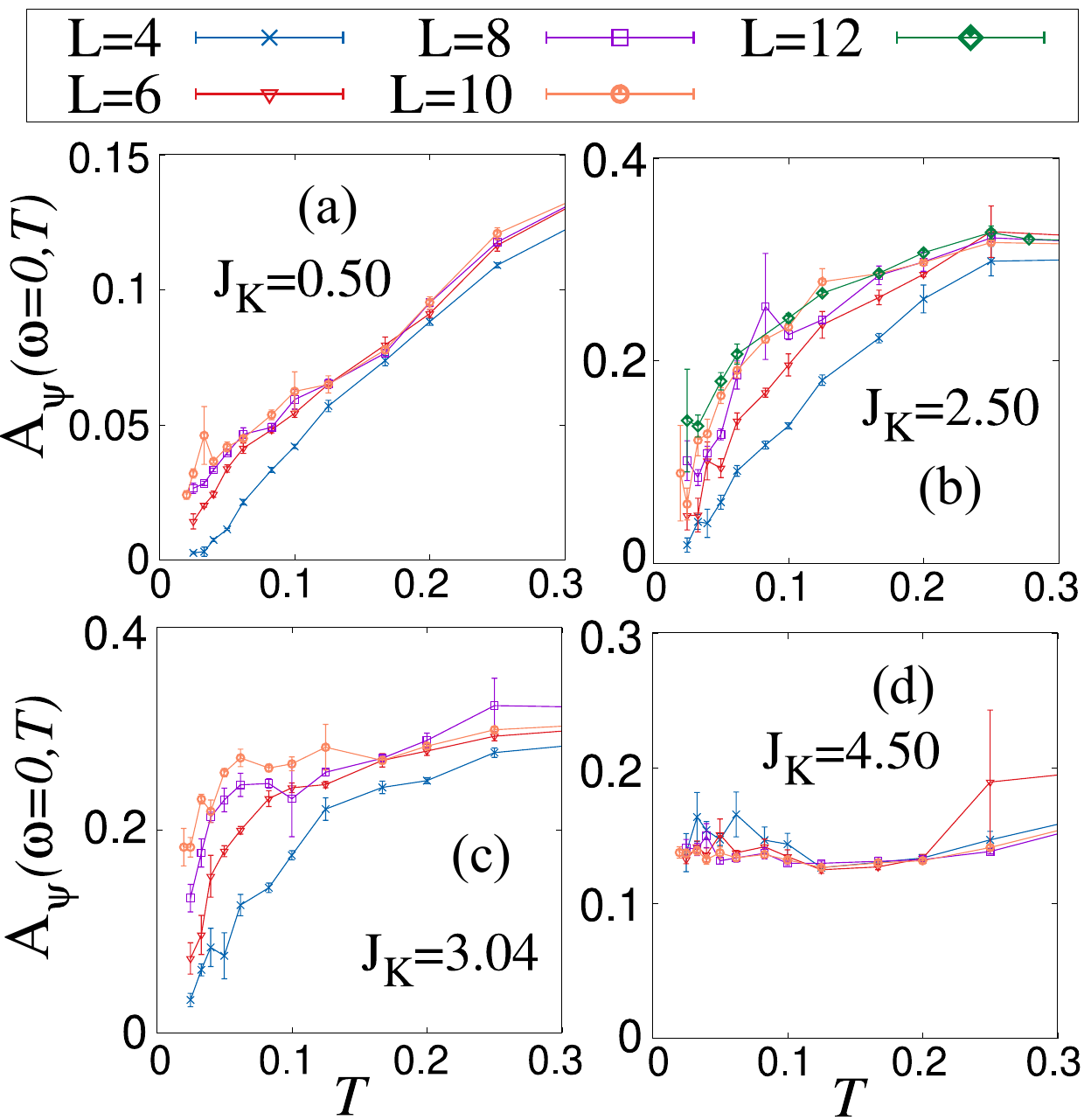}
\par\end{centering}
\caption{Same as Fig.~\ref{fig:cdos_size_effect}, but for the composite-fermion local density of states $A_\psi(\omega=0,T)$.}
\label{fig:fdos_size_effect}
\end{figure}

In this appendix, we provide additional plots for the local density of states $A_{c/\psi}(\omega=0,T)$, in order to illustrate the finite-size effects. In Figs.~\ref{fig:cdos_size_effect} and \ref{fig:fdos_size_effect}, we compare the data obtained from different system sizes, using different values of the Kondo coupling $\Jk$. For the considered range of parameters, $T>0.025$, $\Jk>0.1$, lattices with linear system sizes $L\ge 10$ appear to be representative of the thermodynamic limit.

\section{Spin-wave theory in antiferromagnetic phase}
\label{spin_wave.app}
% !TEX root = 2d_3d_kondo_prb.tex
To confirm the existence of Landau-damped Goldstone modes in the antiferromagnetic phase within a spin-wave description, we perform a Holstein-Primakoff transformation of the local moments. The latter are then perturbatively coupled to the conduction electrons. 
From a theoretical perspective, this procedure provides a combined expansion
in both the inverse spin lengths of the local moments $1/S$ and the Kondo coupling $\Jk$. 
In particular, if the magnon modes can dissipate energy by exciting electrons, this is reflected in the magnon propagator, which we will compute in the following. 

\begin{figure}[tb]
\begin{centering}
\includegraphics[width=0.48\textwidth]{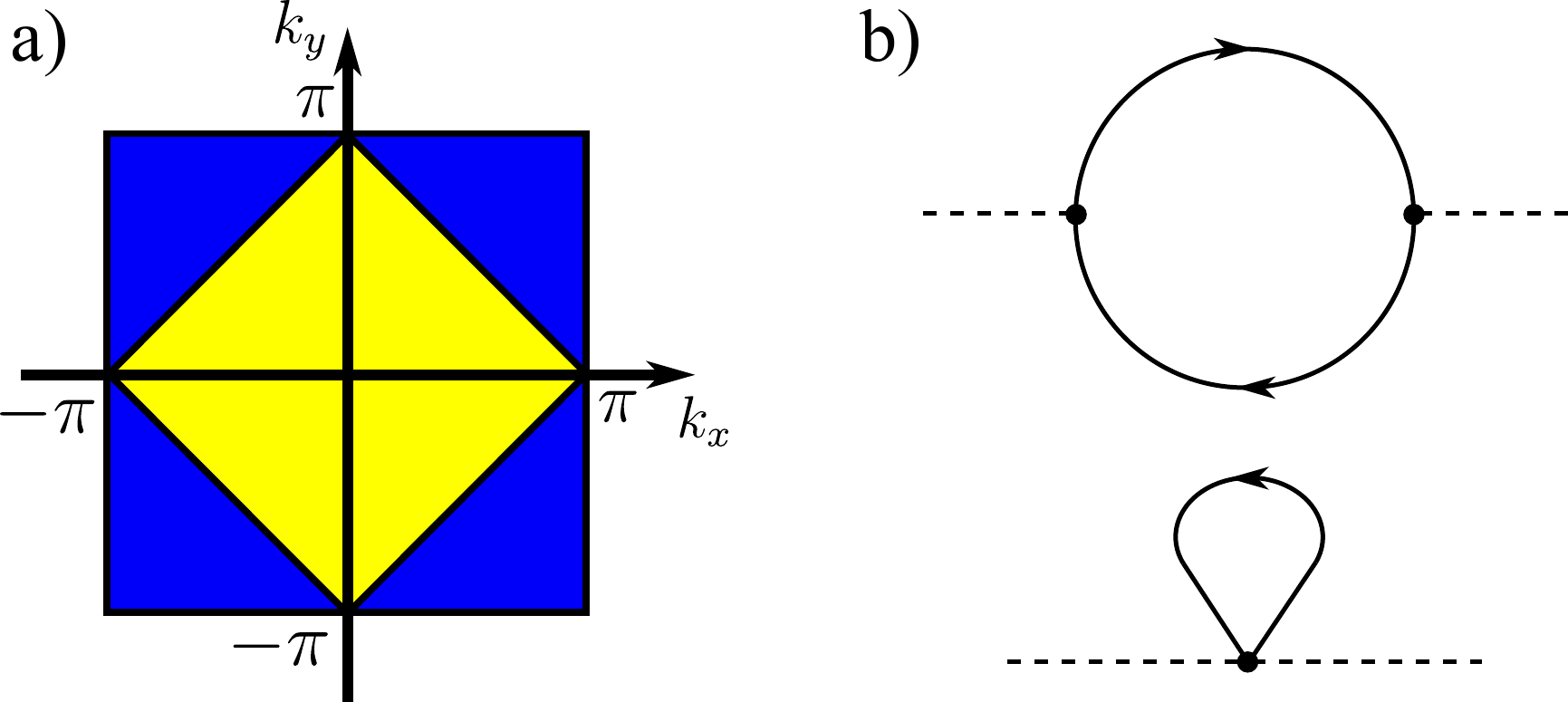}
\par\end{centering}
\caption{(a) Crystallographic Brillouin zone (blue) and backfolded magnetic Brilliouin zone (yellow). (b) Magnon self-energies $\pi$ in one-loop perturbation theory from Eqs.~\eqref{eq:Hlin} (top) and \eqref{eq:Hquad} (bottom).}
\label{fig:magBZ+Feyn}
\end{figure}

In the absence of Kondo interactions,
the local moments form a
Heisenberg antiferromagnet in the impurity layer.  
We introduce two sublattices $A$ and $B$ to take the staggered magnetization into account. The leading order of the Holstein-Primakoff transformation in the limit $S \to \infty$ reads
\begin{align}\label{eq:HolsPrim}
\begin{split}
\arraycolsep=.5 cm
\begin{array}{ l l}
\hat{{S}}^+_{\ve{i} \in A}   \simeq \sqrt{2S} a_{\ve{i}} &
\hat{{S}}^+_{\ve{i} \in B}   \simeq \sqrt{2S} b^\dagger_{\ve{i}} \\
\hat{{S}}^-_{\ve{i} \in A}   \simeq \sqrt{2S} a^\dagger_{\ve{i}} &
\hat{{S}}^-_{\ve{i} \in B}   \simeq \sqrt{2S} b_{\ve{i}}\\
\hat{{S}}^z_{\ve{i} \in A}  = S - a^\dagger_{\ve{i}} a_{\ve{i}} &
\hat{{S}}^z_{\ve{i} \in B}  = - S + b^\dagger_{\ve{i}} b_{\ve{i}} 
\end{array} \, ,
 \end{split}
\end{align}
where the bosonic operators $a^{(\dagger)}_{\ve{i}},b^{(\dagger)}_{\ve{i}}$ annihilate (create) a spin-wave excitation. To consider the coupling between these magnons and the conduction electrons, we first have to establish a description of the full three-dimensional electronic band structure that is compatible with the N\'eel order in the impurity layer at $R_z=0$. To this end, we formally introduce the same $A,B$ sublattices in all layers with $A-A$ stacking in the $R_z$ direction, such that hopping processes between layers with different $R_z$, but identical in-plane coordinate $\ve i$, do not change the sublattice type. This reformulation is equivalent to considering a square lattice in each layer with a basis that contains two neighboring sites of the original lattice. Like in a two-dimensional system, the corresponding band structure is therefore obtained by backfolding the in-plane part of the dispersion relation $\epsilon_{\ve{k}_2,k_z}$ with $\ve{Q}$, see Fig.~\ref{fig:magBZ+Feyn}(a). This gives rise to the two new bands
\begin{align}
\epsilon^{(1,2)}_{\ve k_2,k_z} & = \mp 2t (\cos k_x + \cos k_y) -2 t \cos k_z = \pm \epsilon_{\ve k_2} + \epsilon_{k_z} \, .
\end{align}
In this way, $\hat{H}_\mathrm{Fermi}$ from Eq.~\eqref{eq:HFermi} becomes
\begin{align}
\hat{H}_\mathrm{Fermi}  =\sum_{\substack{\boldsymbol{k}_{2},k_{z},\sigma \\ n=1,2}}\epsilon^{(n)}_{\boldsymbol{k}_{2},k_{z}}\hat{c}_{\boldsymbol{k}_{2},k_{z},n,\sigma}^{\dagger}\hat{c}_{\boldsymbol{k}_{2},k_{z},n,\sigma}^{\phantom{\dagger}}.
\end{align}
and consequently, also the associated bare propagator acquires a matrix form in the band space with indices $n_{1,2}=1,2$
\begin{align}\label{eq:G0n1n2}
 (\underbar G^{(0)})_{n_1,n_2,k_z,q_z,\sigma}(\ve k_2,\omega) = 
 \left(
 \begin{array}{c c}
  \frac{\delta_{k_z q_z}}{i \omega - \epsilon^{(1)}_{\ve k_2,k_z}} & \underline 0 \\
  \underline 0 & \frac{\delta_{k_z q_z}}{i \omega - \epsilon^{(2)}_{\ve k_2,k_z}}
 \end{array}
 \right) \, ,
\end{align}  
where the matrix structure in the out-of-plane momentum space has been introduced, anticipating the violation of momentum conservation by the interactions. In analogy to the main text, we define the local propagators at $R_z=0$, $g^{(0)}_{n_1,n_2,\sigma}(\ve k_2,\omega) = 1/L \sum_{k_z,q_z} (\underbar G^{(0)})_{n_1,n_2,k_z,q_z,\sigma}(\ve k_2,\omega) $. For the tight-binding dispersion considered here, one has 
\begin{align}\label{eq:gloc0}
g^{(0)}_{n_1,n_2,\sigma}(\ve k_2,\omega) = \frac{\delta_{n_1,n_2}}{\sqrt{(i \omega - \epsilon_{\ve k_2} +2t)} \sqrt{(i \omega - \epsilon_{\ve k_2} -2t)}} \, . 
\end{align}
Next, we turn to the perturbative corrections arising from $\hat H_{\rm{Kondo}}$ given in Eq.~\eqref{eq:HKondo}. 
The most important term is of order $\mathcal{O}(S \Jk)$ and describes the interaction with the static staggered magnetization contained in the $\hat{S}^z$ components of Eq.~\eqref{eq:HolsPrim}. The corresponding mean-field-like Hamiltonian reads
\begin{align}\label{eq:HMF}
 H_\text{MF} =- \frac{\Jk S}{2 L} \sum_{\ve k_2,k_z,q_z,\sigma} \sigma \left(c^\dagger_{\ve k_2,k_z,\sigma,1} c_{\ve k_2,q_z,\sigma,2}+ \text{h.c.}\right)\, ,
\end{align}
where spin up (down) correspond to the value $\sigma=\pm 1$. Since the perturbation is quadratic, it gives rise to the static self-energy $(\underline\Sigma^\mf)_{n_1,n_2,k_z,q_z,\sigma} = \tilde\Sigma^\mf_{n_1,n_2,k_z,q_z,\sigma}/ L = - \sigma \Jk S (1-\delta_{n_1,n_2}) /(2L)$, which is independent of momentum. 
The fact that the self-energy is nonzero only for interband processes stems from the staggered magnetization: Any scattering event from the alternating pattern translates to a shift by $\ve{Q}$ in momentum space that connects identical wave vectors, but changes the band index. 
Note that $\underline \Sigma^\mf$ is equivalent to the mean-field form discussed above Eq.~\eqref{eq:AcMF}, with the additional simplification that in the perturbative regime the staggered magnetization is given by $m^f = S$. 
The solution to the Dyson equation $ [\underline G^\mf]^{-1} = \underline [G^{(0)}]^{-1}-\underline\Sigma^\mf$ is given in terms of the scattering form
\begin{align}\label{eq:resGMF}
\begin{split}
 &\underline G_\sigma^\mf(\ve k_2,\omega) = \\
  & \qquad\underline G_\sigma^{(0)}(\ve k_2,\omega) +  \underline G_\sigma^{(0)}(\omega,\ve  k_2) \cdot \underline T_\sigma^\mf(\ve k_2,\omega) \cdot \underline G_\sigma^{(0)}(\ve k_2,\omega) \,,
  \end{split}
  \end{align}
  where the dots denote matrix multiplication both in band and $k_z$ space, and the $T$ matrix is given at the mean-field level by
  \begin{align}\label{eq:TMF}
  \begin{split}
& \underline T_\sigma^\mf(\ve k_2,\omega) = \frac{1/L}{1-(S \Jk/2)^2 g^{(0)}_{11,\sigma}(\ve k_2,\omega) g_{22,\sigma}^{(0)}(\ve k_2,\omega)} \\ 
 & \times \left(
 \begin{array}{c c}
  (S \Jk/2)^2 g^{(0)}_{22,\sigma}(\ve k_2,\omega)  \underline A & -\sigma S \Jk/2 \cdot \underline A \\
  -\sigma S \Jk/2 \cdot \underline A & (S \Jk/2)^2 g^{(0)}_{11,\sigma}(\ve k_2,\omega) \underline{A}
 \end{array}
\right)\, .
\end{split}
\end{align}
As in Eq.~\eqref{eq:G0n1n2}, the outer $2 \times 2$ matrix refers to the band index, whereas the out-of-plane momentum structure for $L$ layers is incorporated by the inner $L \times L$ matrices proportional to $\underline A$. Since the Kondo interaction is localized in only of the layers, $ H_\text{MF}$ does not introduce correlations between the initial and final out-of-plane momenta, and we have $(\underline A)_{k_z,q_z} = 1$ for all $k_z, q_z$.
Physically, $\underline T^\mf_\sigma$ can be understood as the $T$ matrix that arises from scattering a single particle off a $\delta$ potential of strength $-\sigma S \Jk$. However, even (odd) powers of $\Jk$ appear in the diagonal (off-diagonal) terms, because a single scattering event changes the band index. Note that $\underline G_\sigma^\mf$ includes all orders of $S$ and $\Jk$, which turns out crucial in order to perform a consistent expansion.

Next, we have to consider the perturbative interaction terms from $H_{\rm{Kondo}}$ that contain magnon fields. From the Holstein-Primakoff transformation~\eqref{eq:HolsPrim} of the spin components $S^{x,y}_{\ve{i}}$, one obtains a Hamiltonian that is linear in the magnons and involves spin flips in the electron sector,
\begin{align}\label{eq:Hlin}
\begin{split}
 \Hlin\! =  \frac{\Jk \sqrt{S}}{2 L} \!\!\! & \sum_{\substack{\ve p_2, \ve q_2\\n=1,2}}\!\!\!  \left[(c^\dagger_{\ve q_2-\ve p_2, R_z=0, \uparrow, n } c_{\ve q_2, R_z=0, \downarrow, n})(a^\dagger_{\ve p_2} +b_{-\ve p_2}) \right. \\
 & \left. +(c^\dagger_{\ve q_2-\ve p_2, R_z=0, \uparrow, n } c_{\ve q_2, R_z=0, \downarrow, \bar n})(b_{-\ve p_2}-a^\dagger_{\ve p_2})+\rm{h.c.}\right]\!.
\end{split}
\end{align}   
Here, $\bar n$ denotes the complementary value of $n$ in band space, i.e., if $n=1$, then $\bar n=2$, and vice versa. In other words, the first line describes intraband and the second line interband processes.   
In addition, we have the two-magnon terms from the $\mathcal O(S^0)$ contribution to $S^z_{\ve i}$ in Eq.~\eqref{eq:HolsPrim}:
\begin{align}\label{eq:Hquad}
\begin{split}
 \Hquad \! = \! \frac{\sigma \Jk}{2L^2} \!\!\! \!\!& \sum_{\substack{\ve q_2,\ve p_2,\ve l_2\\ \sigma,n=1,2}}\!\!\!\!\!   \left[\!(c^\dagger_{\ve q_2+ \ve l_2-\ve p_2,R_z=0,\sigma,n} c_{\ve q_2,R_z=0,\sigma,n}\!) (b^\dagger_{\ve p_2} b_{\ve l_2}\!\!-\!a^\dagger_{\ve p_2} a_{\ve l_2}) \right.\\
& \left. +(c^\dagger_{\ve q_2+ \ve l_2-\ve p_2,R_z=0,\sigma,n} c_{\ve q_2,R_z=0,\sigma,\bar n}) (a^\dagger_{\ve p_2} a_{\ve l_2} + b^\dagger_{\ve p_2} b_{\ve l_2})\right].
\end{split}
 \end{align}
To incorporate the effects of $H_{1,2}$ properly via perturbation theory, we use a coherent-state path integral formulation and integrate out the conduction electrons. This yields the effective partition function $\mathcal Z = \int \mathcal{D}[a,b] \exp(- S[a,b])$ of the magnons. 
In particular, the quadratic part of the action $S[a,b]$ acquires self-energy corrections by the Kondo interactions but the condition $\<a_{\ve p_2}\> = 0=\<b_{\ve p_2}\>$ is retained since the mean-field expectation value of the staggered magnetization remains unchanged. The dressed quadratic magnon action reads
\begin{align}\label{eq:Smag}
S[a,b] = \int \frac{d\Omega}{2\pi} \sum_{\ve p_2}
\left(
\begin{array}{c}
	a^\ast_{\ve p_2, \Omega} \\
	b_{-\ve p_2, -\Omega}
\end{array}
\right)
\underline D^{-1}(\Omega, \ve p_2)
\left(
\begin{array}{c}
	a_{\ve p_2, \Omega} \\
	b^\ast_{-\ve p_2, -\Omega}
\end{array}
\right),
\end{align}
where
\begin{align}
\underline{D}^{-1}=\left(
\begin{array}{c c}
	- i\Omega +\mathcal E_{\mathrm{H}} -\pi_{a^\ast a}(\ve p_2, \Omega) &  \mathcal E_{\mathrm{H}} \gamma_{\ve p_2}^\ast - \pi_{a^\ast b^\ast}(\ve p_2, \Omega) \\  \mathcal E_{\mathrm{H}} \gamma_{\ve p_2} - \pi_{b a}(\ve p_2, \Omega) & i\Omega +\mathcal E_{\mathrm{H}} -\pi_{bb^\ast}(\ve p_2, \Omega)
\end{array}
\right).
\end{align}
Here, the contributions at vanishing $\Jk$ that stem from the Heisenberg Hamiltonian are encoded in $\mathcal E_{\mathrm{H}} = \Jh S z_\text{c}$, with coordination number $z_\text{c}$ and $\gamma_{\ve p_2} = 2(\cos p_x +\cos p_y)/z_\text{c}$. In contrast, the $\pi_{ab}$ functions include the effects from the Kondo interactions.  
The lowest-order self-energies generated by $H_1$ and $H_2$ are depicted in Fig.~\ref{fig:magBZ+Feyn}(b), in which the internal electron lines are to be evaluated using $\underline G^\mf$. 

For the remaining part of this appendix, we assume the limits of large $S$ and small $\Jk$ in a way that the product $\Jk S$ remains small. As a result, the the particle-hole bubbles, which are generated by $\Hlin^2 \sim \Jk^2 S$, can be evaluated with $\underline G^{(0)}$,  
since the next-to-leading term from the $T$ matrix in $\underline G^\mf$ is suppressed by an additional factor $\Jk S$. Therefore, the corresponding self-energies result in noninteracting local density-density correlation functions,
\begin{align}\label{eq:pilin}
\begin{split}
&\pi^{(1)}_{a^\ast a}(\ve p_2, \Omega)  = -\frac{\Jk^2 S}{4} \sum_{n_1,n_2=1,2} \Pi^{(0)}_{n_1,n_2}(\ve p_2, \Omega)\,, \\
&\pi^{(1)}_{a^\ast b^\ast}(\ve p_2, \Omega)   = -\frac{\Jk^2 S}{4} \sum_{n_1,n_2=1,2} (-1)^{n_1-n_2} \Pi^{(0)}_{n_1,n_2}(\ve p_2, \Omega) \, ,
\end{split}
\end{align} 
and furthermore we have
$\pi^{(1)}_{b b^\ast}(\ve p_2, \Omega) = \pi^{(1)}_{a^\ast a}(\ve p_2, \Omega)$ and $\pi^{(1)}_{b a}(\ve p_2, \Omega) = \pi^{(1)}_{a^\ast b^\ast}(\ve p_2, \Omega)$. In the above, the band-selective correlation functions read
\begin{align}\label{eq:defPi0}
\begin{split}
\Pi^{(0)}_{nl}(\ve p_2, \Omega) \! = \!\! \int \!\!\!\frac{d^2 q_2}{(2\pi)^2}\!\! \int\!\!\! \frac{d\omega}{2\pi}  g^{(0)}_{nn\uparrow} (\ve q_2+\ve p_2,\omega+\Omega)
g^{(0)}_{ll\downarrow} (\ve q_2,\omega).
\end{split}
\end{align}
Before evaluating them, we consider the perturbative correction from $H_2$ that gives rise to the fermion loop formed by a single $\underline G^\mf$, yielding the constant
\begin{align}\label{eq:piquad}
\begin{split}
& \pi^{(2)}_{a^\ast a}(\ve p_2, \Omega) = \\
& \frac{ \Jk}{2L^3} \int \frac{d\omega}{(2\pi)} \sum_{\substack{\ve k_2,k_z,q_z\\ \sigma,n_{1},n_2}}   (-1)^{n_1-n_2} \sigma (\underline G^\mf_\sigma(\ve k_2, \omega))_{n_1,n_2,k_z,q_z} \\
& \quad\qquad\qquad \simeq \frac{ \Jk^2 S}{2}\left[\Pi^{(0)}_{12}(\ve 0, 0) + \Pi^{(0)}_{21}(\ve 0, 0) \right]\, ,
\end{split}
\end{align}
where the second line refers to the lowest order in $\Jk$.
Note that the diagonal terms of $\underline G^\mf$ cancel identically by symmetry to all orders, such that one finds analogously $\pi^{(2)}_{a^\ast a}(\ve p_2, \Omega) = \pi^{(2)}_{b^\ast b}(\ve p_2, \Omega)$. In addition, we have $\pi^{(2)}_{b a}(\ve p_2, \Omega) = 0 =\pi^{(2)}_{a^\ast b^\ast}(\ve p_2, \Omega)$.
The total magnon self-energies are obtained by adding the contributions from Eqs.~\eqref{eq:pilin} and~\eqref{eq:piquad}, i.e.,
\begin{align}\label{eq:pifin}
\begin{split}
&\pi_{a^\ast a}(\ve p_2, \Omega)  = -\frac{\Jk^2 S}{4} \sum_{n_1,n_2=1,2} \Pi^{(0)}_{n_1,n_2}(\ve p_2, \Omega)\\
&\qquad\qquad\qquad\quad + \frac{ \Jk^2 S}{2}\left[\Pi^{(0)}_{12}(\ve 0, 0) + \Pi^{(0)}_{21}(\ve 0, 0) \right]\,,  \\
&\pi_{a^\ast b^\ast}(\ve p_2, \Omega)   = -\frac{\Jk^2 S}{4} \sum_{n_1,n_2=1,2} (-1)^{n_1-n_2} \Pi^{(0)}_{n_1,n_2}(\ve p_2, \Omega) \, ,
\end{split}
\end{align}
and $\pi_{a^\ast a}(\ve p_2, \Omega) = \pi_{b^\ast b}(\ve p_2, \Omega)$, $\pi_{b a}(\ve p_2, \Omega) = \pi_{a^\ast b^\ast}(\ve p_2, \Omega)$. 
The above implies the periodicity properties $\pi_{a^\ast a}(\ve p_2+l\ve Q, \Omega) = \pi_{a^\ast,a}(\ve p_2, \Omega)$,
$\pi_{a^\ast b^\ast}(\ve p_2 + l \ve Q, \Omega) = (-1)^l \pi_{a^\ast b^\ast}(\ve p_2+ l\ve Q, \Omega) $ for $l \in \mathbb{Z}$.
Moreover, all the magnon self-energies approach, in the limit $\Omega \to 0, \ve p_2 \to \ve 0$, the same value,
\begin{align}\label{eq:pires}
-\frac{\Jk^2 S}{4}\left[\Pi^{(0)}_{12}(\ve 0, 0)+\Pi^{(0)}_{12}(\ve 0, 0)-\Pi^{(0)}_{12}(\ve 0, 0) - \Pi^{(0)}_{21}(\ve 0, 0) \right] \, .
\end{align}
Next, we diagonalize $S[a,b]$ from Eq.~\eqref{eq:Smag} via a bosonic Bogoliubov transformation in the presence of the magnon self-energies,
\begin{align}\label{eq:BogT}
\left(
\begin{array}{c}
a_{\ve p_2, \Omega} \\
b^\ast_{-\ve p_2, -\Omega}
\end{array}
\right)
=
\left(
\begin{array}{c c}
	u_{\ve p_2,\Omega} & v_{\ve p_2,\Omega} \\
	v_{\ve p_2,\Omega} & u_{\ve p_2,\Omega}
\end{array}
\right)
\left(
\begin{array}{c}
\alpha_{\ve p_2, \Omega} \\
\beta^\ast_{-\ve p_2, -\Omega}
\end{array}
\right) \, .
\end{align}
As usual, the real parameters $u_{\ve p_2,\Omega}, v_{\ve p_2,\Omega}$ satisfy $u_{\ve p_2,\Omega}^2- v_{\ve p_2,\Omega}^2=1$ to keep the measure of the path integral invariant. The standard parametrization
$u_{\ve p_2,\Omega} = \cosh \theta_{\ve p_2,\Omega}$, $v_{\ve p_2,\Omega} = \sinh \theta_{\ve p_2,\Omega} $ and the choice $\tanh(2\theta_{\ve p_2,\Omega}) = - [\mathcal{E}_\mathrm{H} \gamma_{\ve p_2}^\ast - \pi_{a^\ast b^\ast}(\ve p_2, \Omega)]/[\mathcal{E}_\mathrm{H} - \pi_{a^\ast a}(\ve p_2, \Omega)]$
yield the diagonal action 
\begin{align}\label{eq:SmagDiag}
S[\alpha,\beta]  =  \int \frac{d\Omega}{(2\pi)}\sum_{\ve k_2}  
\left(
\begin{array}{c}
\alpha^\ast_{\ve p_2, \Omega} \\
\beta_{-\ve p_2, -\Omega}
\end{array}
\right) 
\tilde{\underline D}^{-1}(\ve p_2, \Omega)
\left(
\begin{array}{c}
\alpha_{\ve p_2, \Omega} \\
\beta^\ast_{-\ve p_2, -\Omega}
\end{array}
\right),
\end{align}
with
\begin{align}
\tilde{\underline D}^{-1}= 
\left(
\begin{array}{c c}
	-i\Omega +\varepsilon(\ve p_2, \Omega) & 0 \\
	0 & i\Omega +\varepsilon(\ve p_2, \Omega)
\end{array}
\right)
\end{align}
and the frequency-dependent dispersion
\begin{align}
\begin{split}
\varepsilon(\ve p_2, \Omega) & \! = \! \sqrt{(\mathcal{E}_{\mathrm{H}}\! -\! \pi_{a^\ast a}(\ve p_2, \Omega))^2-(\mathcal{E}_{\mathrm{H}} \gamma_{\ve p_2}^\ast \! \!- \! \pi_{a^\ast b^\ast}(\ve p_2, \Omega))^2} .
\end{split}
\end{align}
Let us consider the limit of low frequencies and momenta. With $\gamma_{\ve{p}_2 \to l \ve Q} = (\cos p_x +  \cos p_y )/2 \approx (-1)^l[1 -  (\ve p_2-l\ve Q)^2/4]$ for coordination number $z_\text{c}=4$, 
the periodicity properties given below Eq.~\eqref{eq:pifin}, and the common value from Eq.~\eqref{eq:pires}, we find that $\varepsilon(0,l\ve Q)$ vanishes for all integer $l$, such that the magnons are gapless at small energies, as expected from Goldstone's theorem. At small, but finite, frequency $\Omega$ and momentum deviations $\delta \ve p_2 = \ve p_2- l \ve Q$, one has the dispersion including only the leading contributions in $\Jk$,
\begin{multline}\label{eq:magnondisp1}
\varepsilon(\delta \ve p_2, \Omega) \\
\simeq \sqrt{\tilde c_B^2 (\delta \ve p_2)^2 - 2 [\pi_{a^\ast a}(\delta\ve p_2, \Omega)- \pi_{a^\ast b^\ast}(\delta\ve p_2, \Omega)]\mathcal{E}_\textrm{H}}\,,
\end{multline}
with the renormalized magnon speed $\tilde c_B = \frac12 \sqrt{\mathcal{E}^2_\mathrm{H} - 2\mathcal{E}_\mathrm{H} \pi(\ve 0, 0)}$. The leading asymptotic behavior of $\pi_{a^\ast a}(\delta\ve p_2, \Omega)- \pi_{a^\ast b^\ast}(\delta\ve p_2, \Omega)$ stems from the discontinuity of the local propagators from Eq.~\eqref{eq:gloc0} $\Im g_{n_1,n_2,\sigma}(\omega \to 0,\ve k_2) \to \delta_{n_1,n_2} \theta(2t-|\epsilon_{\ve k_2}|)\text{sgn}(\omega)/(4t^2-\epsilon_{\ve k_2}^2) $, which is finite only in the projected two-dimensional Fermi surface shown in Fig.~\ref{fig:plot_of_model}(c). As a result, we obtain the nonanalytic behavior  
\begin{align}\label{eq:deltaPi}
\begin{split}
& \pi_{a^\ast a}(\delta\ve p_2, \Omega)- \pi_{a^\ast b^\ast}(\delta\ve p_2, \Omega) \\ 
& \simeq - \Jk^2 S \int_{\left[\substack{\text{proj.}\\ \text{2D FS}}\right]'} \frac{d^2 k_2}{(2\pi)^2} \int \frac{d\omega}{2\pi} \frac{\text{sgn}(\omega+\Omega) \text{sgn}(\omega)-\text{sgn}(\omega)^2}{4t^2-\epsilon^2_{\ve k_2}} \\
&= \frac{2J^2_\mathrm{K} S}{\pi} |\Omega| \int_{\left[\substack{\text{proj.}\\ \text{2D FS}}\right]'} \frac{d^2 k_2}{(2\pi)^2} \frac{1}{4t^2-\epsilon^2_{\ve k_2}} \equiv \Jk^2 S \alpha |\Omega|\, .
\end{split}
\end{align}
Note that the van-Hove singularities at the boundary of the projected two-dimensional Fermi surface are regularized by inserting the full $\underline G^\mf$ rather than only the noninteracting part, such that $\alpha$ is well-defined. In the above this is indicated by the prime on the integration boundary. As a consequence, the low-frequency asymptotics of the  magnon dispersion,
\begin{align}
\varepsilon(\delta \ve p_2 \to 0, \Omega \to 0) \simeq \sqrt{\tilde c^2_B(\delta \ve p_2^2)^2 + \mathcal{E}_\mathrm{H} \Jk^2 S \alpha|\Omega|}\, ,
\end{align} 
is dressed by a frequency-dependent term that has the same functional form as Landau damping (in imaginary frequencies), generated by the density fluctuations of a Fermi gas~\cite{Altland2010} at finite wave vector $\ve Q$. These results allow to make contact with the scaling analysis of Sec.~\ref{Weak_coupling:sec} for the weak coupling regime: By Fourier transformation, the Landau damping correction is associated with
the scale-invariant temporal fluctuations at large imaginary times in the dressed action $S^{(0)}$ of Eq.~\eqref{S_landau_damped.eq}. Similarly, the renormalization of the magnon speed is expected to correspond to the long-distance fluctuations at equal times.  

\begin{figure}[tb]
\begin{centering}
\includegraphics[width=\linewidth]{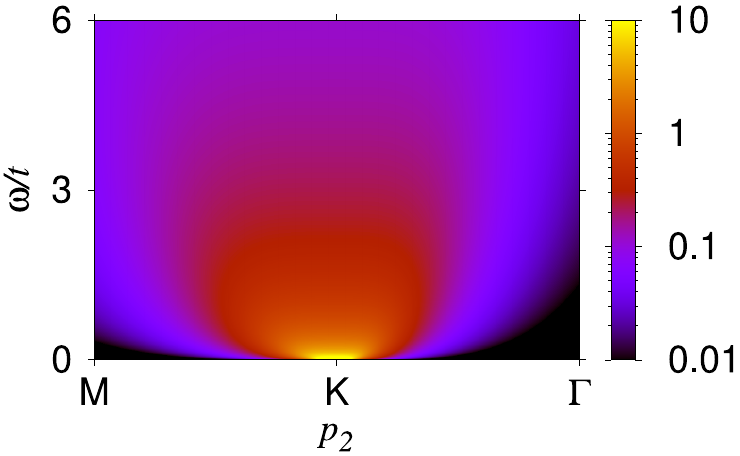}
\par\end{centering}
\caption{Spin spectral function $\chi''(\delta \ve p_2, \omega)$ in Eq.~(\ref{eq:chiMFT}) along the high-symmetry path $\ve{M}(\pi,0)\rightarrow \ve{K}(\pi,\pi)\rightarrow\ve{\Gamma}(0,0)$. Here, we have used the parameters $t=1$, $\Jk=2.5$, $\Jh=0.5$, $S=1/2$, $\tilde c_B=0.447$, and $\alpha=0.3$.}
\label{fig:chi_HM}
\end{figure}

Finally, we calculate the spin spectral function to connect these results with the numerical simulations discussed in the main text. We start out from the (connected) spin structure factor in imaginary time $S(\tau,\ve i) = -\left\langle \mathcal{T}_{\tau}\left[\hat{\ve S}^f_{\ve i}(\tau) \cdot \hat{\ve S}^f_{\ve i = \ve 0}(0) \right] \right \rangle_c $ where we replace the spin operators via the Holstein-Primakoff transformation~\eqref{eq:HolsPrim} and consider terms up to order $S$. Physically, these correspond to correlations of $\<S^{x,y},S^{x,y}\>$, while fluctuations in the $z$ direction require a change in the magnitude of the staggered magnetization, corresponding to a high-energy process. Next, we Fourier transform to imaginary frequencies and momenta,
followed by the bosonic Bogoliubov transformation~\eqref{eq:BogT}, and obtain:
\begin{align}
S(\ve p_2, \Omega) = - S\left(u^2_{\ve p_2,\Omega} + v^2_{\ve p_2,\Omega} + 2 u_{\ve p_2,\Omega} v_{\ve p_2,\Omega} \right) \frac{2 \varepsilon(\ve p_2,\Omega)}{\Omega^2 +\varepsilon(\ve p_2,\Omega)^2} .
\end{align} 
Inserting the standard identities of bosonic Bogoliubov transforms $u^2_{\ve p_2,\Omega} + v^2_{\ve q_2,\Omega} = \cosh(2 \theta_{\ve p_2,\Omega}) = 1/\sqrt{1-\tanh^2(2 \theta_{\ve p_2,\Omega})}$ and 
$2 u_{\ve p_2,\Omega} v_{\ve p_2,\Omega} = \sinh(2 \theta_{\ve p_2,\Omega}) = \tanh(2 \theta_{\ve p_2,\Omega})/\sqrt{1-\tanh^2(2 \theta_{\ve p_2,\Omega})}$ with the value of $\tanh(2 \theta_{\ve p_2,\Omega})$ given above Eq.~\eqref{eq:SmagDiag}, yields $u^2_{\ve p_2,\Omega} + v^2_{\ve p_2,\Omega} + 2 u_{\ve p_2,\Omega} v_{\ve p_2,\Omega} \to 2 \mathcal E_\mathrm{H} / \varepsilon(\ve p_2, \Omega)$ in the vicinity of the $\ve{K}$ point, $\ve p_2 \to \ve Q$, and small $\Omega $. In total, we find for the structure factor in the limit $\delta \ve p_2 = \ve p_2 - \ve Q \to 0$
\begin{align}
\begin{split}
& S(\delta \ve p_2, \Omega)\! \to \! \frac{-4 \mathcal E_\mathrm{H} S}{\Omega^2 + \varepsilon^2( \delta \ve p_2,\Omega)}
\! = \!  \frac{-4 \mathcal E_\mathrm{H} S}{\Omega^2+\tilde c^2_B \delta \ve p_2^2 + \mathcal E_\mathrm{H} J^2_\mathrm{K} S \alpha|\Omega|} \, .
\end{split}
\end{align}
The spin spectral function $\chi''(\ve p_2, \Omega) = - \pi^{-1} \Im S(i\Omega \to \omega + i 0^+ ,\ve p_2)$ is obtained via analytic continuation. In the vicinity of the $\ve{K}$ point, this results in
\begin{align} \label{eq:chiMFT}
\chi''(\delta \ve p_2, \omega) = \frac{1}{\pi} \frac{4 \mathcal E^2_\mathrm{H} S^2 J^2_K \alpha \omega }{\tilde c_B^2(\delta \ve p_2)^4  +(\mathcal E_\mathrm{H} J^2_K S \alpha \omega)^2} \, ,
\end{align}
where the bare $\omega^2$ term has been neglected, since it is irrelevant for the low-energy asymptotics. In contrast to the pure Heisenberg dynamics with sharp linear magnons at the $\ve K$ point, $\chi''$ exhibits a feature of broadened, Landau-damped magnons in the presence of finite Kondo coupling, see Fig.~\ref{fig:chi_HM}, to be compared with Fig.~\ref{fig:Spin_akw} in the main text. In addition, the scaling of typical frequencies and momenta is given by $\omega \sim \delta \ve p_2^2$, corresponding to the dynamical critical exponent $z=2$, as discussed in Sec.~\ref{Weak_coupling:sec}.

\section{Mean-field theory in paramagnetic phase}
\label{mean_field.app}
% !TEX root = 2d_3d_kondo_prb.tex
In this section, we study the structure of the mean-field theory in further detail, to provide some analytic understanding of the numerical observations made in Sec.~\ref{sec:MF}, in particular, the existence of two-dimensional states. 
We focus on the paramagnetic Kondo phase to simplify the procedure. Setting the magnetic order parameters $m_{c,f}$ and also $\lambda$ to zero, in order to ensure particle-hole symmetry, the mean-field Hamiltonian from Eq.~\eqref{eq:mft_ham} acquires the form
\begin{align}\label{eq:HMFapp}
\hat{H}_{\text{MF}} & = \!\!\!\!\sum_{\substack{\ve k_2, R_z\\ \sigma}}\!\!\! \left[\epsilon_{\ve k_2} c^\dagger_{\ve k_2, R_z,\sigma} c_{\ve k_2, R_z,\sigma}\!-\!t(c^\dagger_{\ve k_2, R_z+1,\sigma} c_{\ve k_2, R_z,\sigma}\! + \!\text{h.c.})\!\right] \nonumber \\ 
& -\Jk V\sum_{\ve k_2,\sigma}\left(\hat{c}_{\ve k_2,R_z=0,\sigma}^{\dagger}\hat{f}_{\ve k_2,\sigma}+\text{h.c.}\right) 
+L^2 \Jk V^2\, .
\end{align} 
Here, we have performed a partial Fourier transform to the in-plane momentum space, but kept the formulation  of the out-of-plane dimension in real space, as in the main text. This brings the in-plane part of the tight-binding Hamiltonian to a diagonal form, as seen in the first line of the above equation. Solving the  mean-field theory requires to diagonalize the given one-particle Hamiltonian. Physically, this is equivalent to finding the eigenstates of the Schr\"odinger equation $\hat H_\mf |\psi\rangle = E |\psi\rangle$. 
For each $\ve k_2$, $\hat H_\mf$ acts on a $L+1$ dimensional Hilbert-space formed by the chain of $L$ $c$  sites along the $z$ direction and the additional $f$ site. In the following, we label the $c$ sites as $R_z=-L/2, -L/2+1,...,0,...L/2-1$ for even $L$ with periodic boundary conditions and the $f$-site simply by $f$. Since only the site at $R_z=0$ is coupled to the additional $f$ orbital via the hybridization parameter $V$, the translational symmetry of the chain is broken in the presence of finite $V$. Consequently, the $L+1$-dimensional eigenvectors $(\psi(R_z),\psi(f))$ of $\hat H_\mf$ are not given by Bloch states.
\begin{figure}[t]
\begin{centering}
\includegraphics[width=\linewidth]{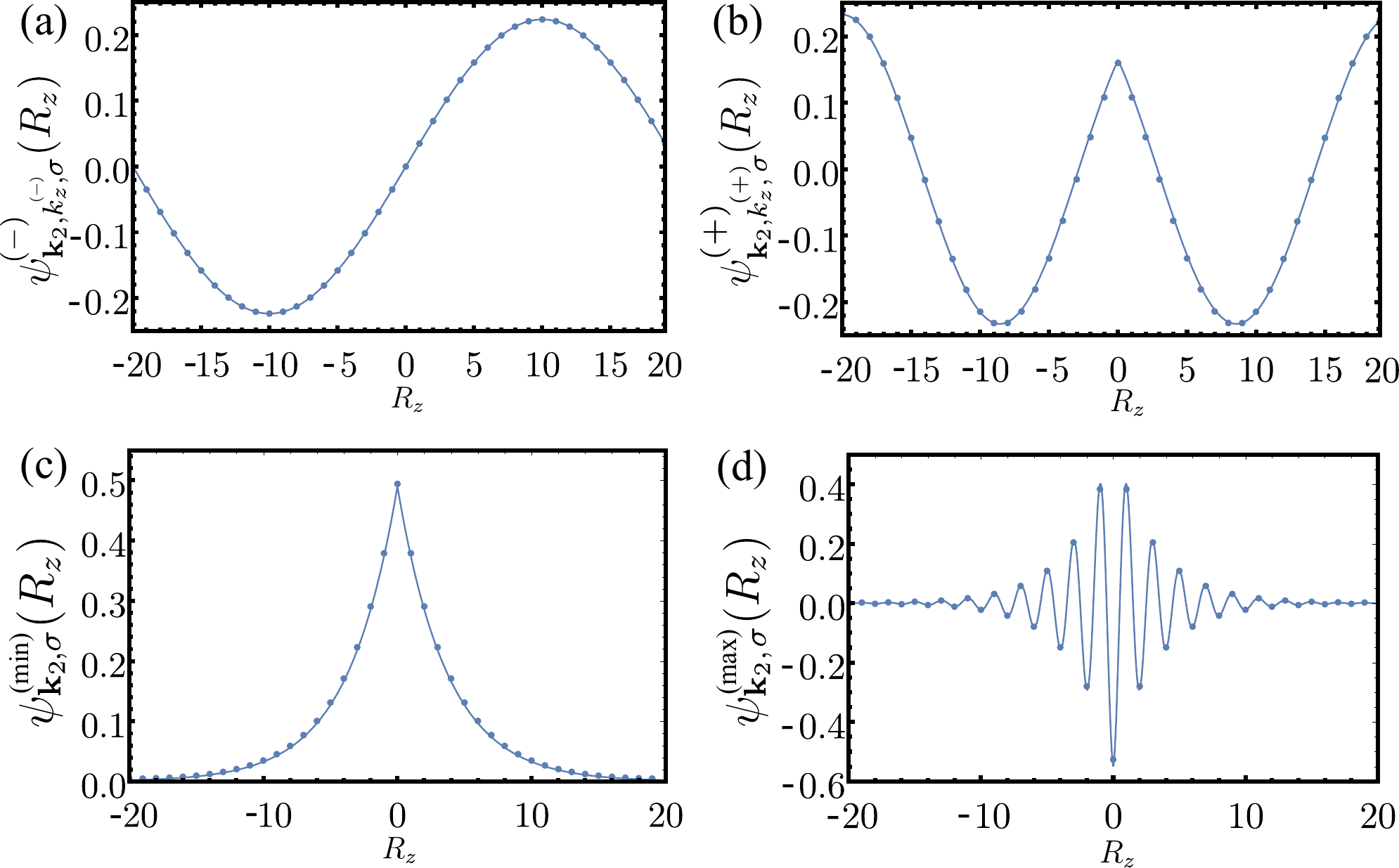}
\par\end{centering}
\caption{Examples of the different wave functions on the $c$ sites for system size $L=40$:
The dots represent the eigenstates obtained by diagonalizing $\hat H_\mf$ numerically at $\Jk/t=5.1$, $V/t = 0.2$, and $\epsilon_{\ve k_2}/t = -0.2$. The solid lines are comparisons with the analytic solutions given in the text for this set of parameters. In particular, the extended states in the first line are (a) odd superpositions of Bloch waves and (b) even scattering states with phase shift. The second line shows the two-dimensional states with (c) minimal and (d) maximal energy.}
\label{fig:wavefunctions}
\end{figure}

There are three different types of wave functions:
\paragraph{Odd superpositions of Bloch waves of the unperturbed lattice at $\mathit{J_K=0}$.} These are characterized by the wave vectors $k^{(-)}_z$,
\begin{align}
\begin{split}
\psi^{(-)}_{\ve k_2, k^{(-)}_z,\sigma}(R_z) & = \sqrt{\frac{2}{L}} \sin( k^{(-)}_z R_z)\,,\\
\psi^{(-)}_{\ve k_2, k^{(-)}_z,\sigma}(f)& =0 \, ,
\end{split}
\end{align}
associated with the noninteracting Bloch energies $E^{(-)}_{\ve k_2, k^{(-)}_z} = \epsilon_{\ve k_2} + \epsilon_{k^{(-)}_z}$ from the tight-binding dispersion. The $k_z^{(-)}$ are given by the wave vectors 
of the free tight-binding chain 
$k_z= 2\pi n/L,\; n \in\{-L/2,-L/2+2, \dots, L/2-1 \}$ excluding $k_z=0$ and $k_z=\pi$, which give only rise to vanishing wave functions. This implies $L/2-1$ different eigenstates  $\psi^{(-)}$. 
Physically, they are not affected by the interactions, because of the zero amplitude in the impurity layer, $\psi^{(-)}_{\ve k_2, k_z,\sigma}(R_z=0)=0$.

\paragraph{Even scattering states.} These are described by a wave vector $k_z^{(+)}$ and an associated phase shift $\phi_{k_z^{(+)}}$:
\begin{align}
\begin{split}
\psi^{(+)}_{\ve k_2, k^{(+)}_z,\sigma}(R_z) & = A^{(+)} \sqrt{\frac{2}{L}}\cos\left( k_z^{(+)} R_z+\phi_{k_z^{(+)}}\sgn R_z \right), \\
\psi^{(+)}_{\ve k_2, k^{(+)}_z,\sigma}(f) & = A^{(+)} \sqrt{\frac{2}{L}} \frac{2\sin k^{(+)}_z \sin \phi_{k_z^{(+)}}}{\Jk V}\, ,
\end{split}
\end{align}
with energy $E^{(+)}_{\ve k_2, k_z} = \epsilon_{\ve k_2} + \epsilon^{(+)}_{k_z}$ and amplitude $A^{(+)}$.
 Note that the wave functions are continuous at $R_z=0$, but the slopes differ when approaching $R_z=0$ from the left or the right. Periodicity restricts the wave vectors to the form 
\begin{align}\label{eq:defkzp}
k_z^{(+)}= \frac{2\pi n -2 \phi_{k_z^{(+)}}}{L} = \frac{k_z -2 \phi_{k_z^{(+)}}}{L}\,,
\end{align} 
while the phase shifts are obtained from the equation
\begin{align}\label{eq:phshift}
\sin k_z^{(+)} \tan \phi_{k_z^{(+)}} = - \frac{\Jk^2 V^2}{2(\epsilon_{\ve k_2}+\epsilon_{k^{(+)}_z})} \, .
\end{align} 
Since the right-hand side is not bounded and may attain both positive and negative values, the phase shifts may vary in the interval $\phi_{k_z^{(+)}} \in [-\pi/2,\pi/2]$. According to Eq.~\eqref{eq:defkzp}, every $k^{(+)}_z$ is therefore adiabatically connected to the closest noninteracting wave vector $k_z$
since adjacent $k_z$ differ by $2\pi/L$. In total, there are $L/2$ different scattering states $\psi^{(+)}_{\ve k_2, k^{(+)}_z,\sigma}$.

\paragraph{Two two-dimensional states.} 
Neglecting corrections that are exponentially small in the system size, we have: 
\begin{enumerate}[(i)]
\item A state below the minimum of the tight-binding dispersion $\min_{k_z}(\epsilon_{\ve k_2} + \epsilon_{k_z})$:  
\begin{align}
\begin{split}
\psi^{(\min)}_{\ve k_2,\sigma}(R_z) & =A^{(\min)} e^{-\kappa |R_z|}\,, \quad \kappa>0 \, , \\
\psi^{(\min)}_{\ve k_2,\sigma}(f) & = A^{(\min)} \frac{2 t \sinh \kappa}{\Jk V}\,,
\end{split}
\end{align}
with energy $E^{(\min)}_{\ve k_2} = \epsilon_{\ve k_2} -2 t \cosh \kappa$. The parameter $\kappa$ satisfies
\begin{align}\label{eq:kappab}
\sinh \kappa = -\frac{\Jk^2 V^2}{2t(\epsilon_{\ve k_2} -2 t \cosh \kappa)}\,,
\end{align}
and the normalization reads $A^{(\min)} = [1-2/(1-\exp(2\kappa))+(2t/\Jk V)^2 \sinh^2 \kappa]^{-1/2}$. Since $\cosh \kappa > 1$ for all real $\kappa \neq 0$, this state has indeed an energy below the tight-binding energies of the extended states discussed before. By inserting $E^{(\min)}_{\ve k_2}$ into the right-hand-side of the last equation, we find furthermore $E^{(\min)}_{\ve k_2} < 0$. 

\item A state of maximal energy above $\max_{k_z}(\epsilon_{\ve k_2} + \epsilon_{k_z})$
\begin{align}
\begin{split}
\psi^{(\max)}_{\ve k_2,\sigma}(R_z) & =A^{(\max)} \cos(\pi R_z) e^{-\kappa |R_z|}\,,\quad \kappa >0\, , \\
\psi^{(\max)}_{\ve k_2,\sigma}(f) & = -A^{(\max)}\frac{2 t \sinh(\kappa) }{\Jk V}\,,
\end{split}
\end{align}
with energy $ E^{(\max)}_{\ve k_2} = \epsilon_{\ve k_2} +2 t \cosh \kappa$, where 
\begin{align}
\sinh \kappa = \frac{\Jk^2 V^2}{2t(\epsilon_{\ve k_2} +2 t \cosh \kappa)}\,,
\end{align}
and $A^{(\max)}=A^{(\min)}$. Here, we have in addition $ E^{(\max)}_{\ve k_2}$.
\end{enumerate}
The presence of two-dimensional states has already been anticipated in Sec.~\ref{sec:MF}: The
spectral functions in Fig.~\ref{fig:Mft_akw} show sharp features of definite sign above and below the continuum of tight-binding energies. Note that these are observable irrespective of the presence of antiferromagnetic order. 

In total, we find $(L/2-1)+L/2+2 = L+1$ eigenstates, as expected for a Hamiltonian of dimension $(L+1)^2$. 
Plots of the different states can be found in Fig.~\ref{fig:wavefunctions}. 

After solving the effective Schr\"odinger equation, we can rewrite the mean-field Hamiltonian from Eq.~\eqref{eq:HMF} in diagonal form as follows
\begin{align}
\begin{split}
 H_\mf  & = \Jk L^2 V^2 + \sum_{\ve k_2 ,\sigma} \sum_{k_z, s=\pm} E^{(s)}_{\ve k_2,k^{(s)}_z} c^\dagger_{\ve k_2,k^{(s)}_z,\sigma} c_{\ve k_2,k^{(s)}_z,\sigma}
\\ & \quad
+\sum_{\ve k_2 ,\sigma} \sum_{s= \min,\max} E^{(s)}_{\ve k_2} c^\dagger_{\ve k_2,s,\sigma} c_{\ve k_2,s,\sigma}\, .
\end{split}
\end{align} 
Here, the new fermionic operators $c_{\ve k_2,k^{(\pm)}_z,\sigma}$ annihilate a quasiparticle in the extended state $\psi^{(\pm)}_{\ve k_2,k^{(\pm)}_z,\sigma}$, whereas $c_{\ve k_2,s,\sigma}$, with $s= \min,\max$, annihilates a fermion in the two-dimensional state $\psi^{(s)}_{\ve k_2,s,\sigma}$, and  analogously for the creation operators. Furthermore, the matrix elements required to transform the operators from the $(R_z,f)$ basis to the new one are also given by the corresponding wave functions. 
In the ground state, the quasiparticles form, at the mean-field level, a Fermi sea with energy
\begin{align}
E_0 =  \Jk L^2 V^2 + \!\sum_{\ve k_2 ,\sigma}\! \frac{1}{2}\sum_{k^{(\pm)}_z}\theta(-E^{(\pm)}_{\ve k_2,k^{(\pm)}_z}) E^{(\pm)}_{\ve k_2,k^{(\pm)}_z}
+ \!\sum_{\ve k_2 ,\sigma}\! E^{(\min)}_{\ve k_2} \, .
\end{align}
The factor $1/2$ in the first sum is needed to avoid double counting of the extended states. 
The mean-field parameter $V$ satisfies the equation $\partial E_0/\partial V =0$. However, one may ask about the behavior of $V$ in the thermodynamic limit $L \to \infty$, since the contributions from the extended states in $E_0$ scale like the volume of the system $L^3$, whereas all other terms only scale like the area of the layer $L^2$. 
To answer this question, we first note that the energies $E^{(-)}$ drop out in $\partial E_0/\partial V$ because they are independent of $V$. 
To simplify the contribution from the extended scattering states with energies $E^{(+)}$, we use the adiabatic connection between them and the Bloch states. We can consider Eqs.~\eqref{eq:phshift} and~\eqref{eq:defkzp} as an iterative scheme to determine both  $k_z^{(+)}$ and $\phi_{k_z^{(+)}}$ at finite $\Jk$. Take as initial condition 
a Bloch state that is part of the noninteracting three-dimensional Fermi surface shown in Fig.~\ref{fig:plot_of_model}(b) with momentum $(\ve k_2, k_z)$, with $k_z \neq 0,-\pi$,  and eigenenergy $\epsilon_{\ve k_2} + \epsilon_{k_z}<0$. 
In this case, Eq.~\eqref{eq:phshift} generates a phase shift that obeys  $\sgn \phi_{k^{(+)}_z} = \sgn k_z$.
As a result, Eq.~\eqref{eq:defkzp} implies $|k_z^{(+)}| < |k_z|$ and therefore $E^{(+)}_{\ve k_2,k^{(+)}_z} < 0$, such that the resulting scattering state at finite $\Jk$ is part of the Fermi sea of quasiparticles. On the contrary, initializing the procedure with a Bloch state with positive energy gives rise to an interacting state with $E^{(+)}_{\ve k_2,k^{(+)}_z} > 0 $. 
Finally, Eqs.~\eqref{eq:phshift} and~\eqref{eq:defkzp} imply that the values $k_z=0$ and $k_z = -\pi$ are mapped to $k^{(+)}_z = \mathcal O(L^{-1})$ and $k^{(+)}_z = -\pi+\mathcal O(L^{-1})$, respectively.  
Consequently, we have $\theta(-E^{(+)}_{\ve k_2,k_z}) = \theta(-\epsilon_{\ve{k}_2} - \epsilon_{k_z})$ in the limit $L \to \infty$ and, moreover, 
\begin{multline}
\sum_{k^{(+)}_z}\theta(-E^{(+)}_{\ve k_2,k^{(+)}_z}) E^{(+)}_{\ve k_2,k^{(+)}_z}   = \\
\sum_{ k_z}\theta(-\epsilon_{\ve k_2} - \epsilon_{k_z}) \left(\epsilon_{\ve k_2} -\epsilon_{k_z} - \frac{2t}{L} \sin k_z \phi_{k^{(+)}_z}+\mathcal O(L^{-2})\right)\, .
\end{multline}
With the above, the mean-field equation for $\partial E_0/\partial V= 0$ becomes independent of system size in the thermodynamic limit,
\begin{multline}\label{eq:MFforV}
\Jk V -t \!\!\int_{\text{3D FS}}\!\! \frac{d^2 \ve k_2 dk_z}{(2\pi)^3} \sin k_z \frac{\partial \phi_{k^{(+)}_z}}{\partial V} = 
2 t \!\! \int \!\! \frac{d^2 \ve k_2}{(2\pi)^2} \frac{\partial \cosh \kappa}{\partial V}\, .
\end{multline} 
The solution to the above equation corresponds to the  behavior of $V$ in the paramagnetic heavy-fermion phase in Fig.~\ref{fig:Mft_phase_diagram}. The asymptotics of $V$ in the limit $\Jk \to \infty$ can be extracted in closed form: Equation \eqref{eq:phshift} entails that all phase shifts approach $\phi_{k^{(+)}_z} \to \pm \pi/2$, irrespective of $V$. Furthermore, from Eq.~\eqref{eq:kappab}, we find $\cosh \kappa \approx \Jk |V|/(2t)$. Searching for a real, positive $V$, the solution approaches
\begin{align}
 \Jk V=2 t \int \frac{d^2 \ve k_2}{(2\pi)^2} \frac{\Jk}{2t} \quad \Rightarrow \quad V= 1 \, ,
\end{align}  
which agrees with the numerical evaluation presented in Sec.~\ref{sec:MF}. Furthermore, we can answer how many quasiparticle states are occupied at a given in-plane momentum $\ve k_2$. As discussed above, the Bloch eigenstates of $\hat{H}_\mathrm{Fermi}$ with wave vector $(\ve k_2,k_z)$ are mapped to the eigenstates of $\hat H_\mf$ as follows: $k_z=0,-\pi$ have corresponding extended scattering states (b) whereas all other pairs $\pm k_z$ form one odd state of type (a) and one even state of type (b). Since $\sgn E^{(\pm)}_{\ve k_2,k_z^{(\pm)}} = \sgn (\epsilon_{\ve k_2} + \epsilon_{k_z})$ we have one quasiparticle for every occupied Bloch state of the noninteracting $3d$ FS.  
In addition, the two-dimensional state (c1) with minimal, negative energy is also occupied. The number of quasiparticles with $\ve k_2$ is thus $\<\hat n_c(\ve k_2)\> +1$. As discussed in the main text, we therefore have $L+1$ electrons per lattice site and the $f$ fermion participates indeed in the Luttinger count.

Finally, we consider the structure of the resulting propagators. The effects of interactions at the mean-field level are most easily determined by representing the hybridization part of $\hat H_\mf$ from Eq.~\eqref{eq:HMFapp} completely in momentum space. This yields the term $-\Jk V L^{-1/2}\sum_{\ve k_2,k_z,\sigma}\left(\hat{c}_{\ve k_2,k_z,\sigma}^{\dagger}\hat{f}_{\ve k_2,\sigma}+\text{h.c.}\right)\, $. Firstly, this implies the self-energy of the conduction electrons
\begin{align}
\Sigma^\mf_c(\omega,\ve k_2)_{k_z, k'_z} = \frac{\Jk^2 V^2}{L} G^{(0)}_f(\omega, \ve k_2) = \frac{\Jk^2 V^2}{L} \frac{1}{i \omega}\, , 
\end{align}
which is off-diagonal in the out-of-plane direction because one-dimensional scattering events break the conservation of momentum. Secondly, the self-energy in the $f$ sector is given by
\begin{align}
\Sigma^\mf_{f}(\omega,\ve k_2) = \Jk^2 V^2 g_{0}(\omega, \ve k_2) \, .
\end{align}
The solution of the Dyson equation for the conduction electrons is therefore obtained by dressing the propagator by the scattering $T$ matrix of the (dynamical) potential,
\begin{multline}\label{eq:resGMFKondo}
G_{c,\sigma}^\mf(\ve k_2,\omega)_{k_z,k'_z} =  G_{c,\sigma}^{(0)}(\ve k_2,\omega,k_z)\delta_{k_z,k'_z}
 \\ 
 +   G_\sigma^{(0)}(\omega,\ve  k_2,k_z) T^\mf(\ve k_2,\omega) G_{c,\sigma}^{(0)}(\ve k_2,\omega, k'_z) \,,
\end{multline}
with the $T$ matrix
\begin{align}\label{eq:TMFKondo}
  \begin{split}
T^\mf(\ve k_2,\omega)_{k_z,k'_z} = \frac{\Sigma_c^\mf(\omega,\ve k_2)_{k_z,k'_z}}{1-\Sigma_c^\mf(\omega,\ve k_2)_{k_z,k'_z} g_{0}(\ve k_2,\omega) }
\, .
\end{split}
\end{align}
We note that Eqs.~\eqref{eq:resGMFKondo} and~\eqref{eq:TMFKondo} correspond to Eqs.~\eqref{eq:resGMF} and~\eqref{eq:TMF}, but here in the case of a single band due to the absence of sublattice magnetization. The propagator of the $f$ electrons reads in turn
\begin{align}
G^\mf_{f} = \frac{1}{i \omega - \Sigma^\mf_f(\omega,\ve k_2)} = \frac{T^\mf(\ve k_2, \omega, R_z=0)}{\Jk^2 V^2} \, ,
\end{align} 
which give rise to the local spectral function in Eq.~\eqref{eq:A_fMF}. After analytic continuation to real frequencies, the two-dimensional-states manifest themselves via poles in the spectral functions $A_{c,f}(\omega,\ve k_2)$ at energies $\omega = \epsilon_{\ve k_2} \pm 2t \cosh \kappa$. As discussed in the main text, the $\ve k_2$-integrated spectral functions show therefore logarithmic van-Hove singularities that are typical for two-dimensional systems.
%%%%%%%%%%%%%%%%%%%%%%%%%%%%%%%%%%%%%%%%%%%%%%%%%%%%%%%%%%%%%%%%%%%%%%%

\bibliographystyle{longapsrev4-2}
\bibliography{2d_3d_kondo_prb,fassaad}

\end{document}